\documentclass[prepreint,twocolumn,trackchanges]{aastex62}
\hypersetup{linkcolor={blue},citecolor={blue}}
\received{\today}
\revised{\today}
\accepted{\today}
\submitjournal{ApJ}

\shorttitle{Ejecta-CSM interaction model for llGRBs}
\shortauthors{Suzuki, Maeda, \& Shigeyama}

\begin{document}
\title{Relativistic supernova ejecta colliding with a circumstellar medium: an application to the low-luminosity GRB 171205A}

\correspondingauthor{Akihiro Suzuki}
\email{akihiro.suzuki@nao.ac.jp}

\author[0000-0002-7043-6112]{Akihiro Suzuki}
\affil{National Astronomical Observatory of Japan, 2-21-1 Osawa, Mitaka, Tokyo 181-8588, Japan}

\author{Keiichi Maeda}
\affiliation{Department of Astronomy, Kyoto University, Kitashirakawa-Oiwake-cho, Sakyo-ku, Kyoto, 606-8502, Japan}

\author{Toshikazu Shigeyama}
\affiliation{Research Center for the Early Universe, School of Science, University of Tokyo, Bunkyo-ku, Tokyo, 113-0033, Japan}



\begin{abstract}
We perform multi-wavelength light curve modeling of the recently discovered low-luminosity gamma-ray burst (GRB) 171205A. 
The emission model is based on the relativistic ejecta-circumstellar medium (CSM) interaction scenario. 
The collision of freely expanding spherical ejecta traveling at mildly relativistic velocities with the CSM produces the reverse and forward shocks, which dissipate a part of the kinetic energy of the mildly relativistic ejecta. 
We show that the early gamma-ray emission followed by an X-ray tail can be well explained by the radiation diffusing out from the shocked gas. 
Mildly relativistic ejecta with a kinetic energy of $5\times10^{50}$ erg and a wind-like CSM with a mass-loss rate of a few $10^{-4}\ M_\odot$ yr$^{-1}$ for a wind velocity of $10^3$ km s$^{-1}$, which extends up to $\sim 3\times 10^{13}$ cm, are required to account for the gamma-ray luminosity and duration of GRB 171205A. 
We also calculate the photospheric and non-thermal emission after the optically thick stage, which can fit the late-time X-ray, optical, and radio light curves. 
Our results suggest that the relativistic ejecta-CSM interaction can be a potential power source for low-luminosity GRBs and other X-ray bright transients. 
\end{abstract}



\section{INTRODUCTION\label{intro}}
Gamma-ray bursts (GRBs) with unusually low gamma-ray luminosities are classified as low-luminosity GRBs (llGRBs) and are thought to comprise a distinct population from their cosmological counterparts. 
Although only a handful of nearby events have been known, their volumetric rate appears to be higher ($10^2$--$10^3$ Gpc$^{-3}$ yr$^{-1}$) than the beaming corrected rate of standard GRBs \citep{2006Natur.442.1011P,2006Natur.442.1014S,2006ApJ...645L.113C,2007ApJ...657L..73G,2007ApJ...662.1111L}. 
They are all associated with broad-lined Ic supernovae (SNe Ic-BL; see \citealt{2006ARA&A..44..507W,2012grbu.book..169H,2017AdAst2017E...5C} for reviews), which exhibit broad-line spectral features. 
Well-observed examples of llGRBs and the associated SNe include, GRB 980425/SN 1998bw \citep{1998Natur.395..663K,1998Natur.395..670G}, GRB 060218/SN 2006aj \citep{2006Natur.442.1008C,2006Natur.442.1018M,2006ApJ...645L..21M,2006Natur.442.1011P,2006Natur.442.1014S}, and GRB 100316D/SN 2010bh \citep{2011MNRAS.411.2792S,2011ApJ...740...41C,2012ApJ...753...67B,2012A&A...539A..76O}.

All the SNe associated with GRBs and llGRBs are suggested to be highly energetic compared with ordinary core-collapse SNe (CCSNe). 
Their light curves and spectra indicate explosion energies $\sim10$ times larger than the canonical value of $10^{51}$ erg. 
The association of GRBs and inferred large kinetic energies imply hidden central engine activities distinguishing these extraordinary events from ordinary explosions of massive stars powered by neutrinos. 
From an observational point of view, scrutinizing the gamma-ray and X-ray emission from these highly energetic CCSNe offers an important tool to investigating their origin. 
The launch of the Neil Gehrels {\it Swift} Observatory \citep{2004ApJ...611.1005G} realized detailed gamma-ray and X-ray observations of GRBs and rapid follow-ups. 
GRBs 060218 and 100316D are well-observed llGRBs detected after the launch of the {\it Swift} satellite. 
They emitted gamma-rays for unusually long priods compared with cosmological GRBs. 
The early X-ray spectra of 060218 and 100316D are well fitted by a power-law spectrum combined with a thermal component with a temperature of the order of $kT\sim 0.1$ keV \citep{2006Natur.442.1008C,2011MNRAS.411.2792S}. 
The long-lasting gamma-ray emission and the large blackbody radii inferred by the thermal component indicate that hot ejecta with a photospheric radius larger than typical radii of compact stars play a vital role in producing high-energy emission.

Radio observations of SNe are another probe for highly energetic explosions through the presence of fast shock waves propagating in the circumstellar medium (CSM) of the exploding star. 
Follow-up radio observations of GRB-SNe have also been conducted. 
GRB-SNe, such as  SN 1998bw \citep{1998Natur.395..663K,2002ARA&A..40..387W}, are indeed known to be a bright radio emitter, which is likely caused by synchrotron emission produced by the forward shock sweeping the ambient gas. 
Furthermore, the discovery of radio-loud SNe Ic-BL without any gamma-ray signature, e.g., SN 2009bb and SN 2012ap, have revealed a population of relativistic SNe, whose radio emission strongly indicates the presence of ejecta traveling at (mildly) relativistic speeds \citep{2010Natur.463..513S,2015ApJ...799...51M,2015ApJ...805..187C}. 
Light curve modelings of radio emission from GRB-SNe and relativistic SNe have also been attempted by several authors \citep[e.g.,][]{2015MNRAS.448..417B,2015ApJ...805..164N}. 

Despite these multi-wavelength observations and intensive discussion, the progenitor system of llGRBs and the origin of their gamma-ray emission are still poorly understood \citep[see, e.g.,][]{2007MNRAS.375..240L,2007ApJ...659.1420T,2007ApJ...664.1026W,2007ApJ...667..351W,2011ApJ...739L..55B}. 
One of the plausible scenarios for the gamma-ray emission is the emergence of a mildly relativistic shock from a CSM in which the progenitor star is embedded (e.g., \citealt{2006Natur.442.1008C,2007ApJ...667..351W,2012ApJ...747...88N}; see also \citealt{1999ApJ...510..379M,1999ApJ...516..788W,2001ApJ...551..946T}). 
In this scenario, the CSM should be sufficiently dense so that the photosphere is well above the surface of the star so as to prolong the prompt gamma-ray emission. 
The mildly relativistic shock can be driven by either a weak jet \citep{2016MNRAS.460.1680I}, the cocoon associated with a jet \citep{2002MNRAS.337.1349R,2005ApJ...629..903L,2013ApJ...764L..12S}, or a jet choked in a star \citep{2011ApJ...739L..55B,2012ApJ...750...68L} or in an extended stellar envelope \citep{2015ApJ...807..172N}.

In the relativistic shock breakout scenario, the forward shock, which propagates in the outermost layer of a star, an extended envelope, or whatever, deposits a fraction of the shock kinetic energy into the internal energy of the shocked gas, which is finally released as bright emission. 
Thus, the shock propagation plays a critical role in how much energy is available for the luminous emission. 
Most theoretical studies on shock breakout scenario \citep[e.g.,][]{2012ApJ...747...88N} consider an accelerating shock in a medium with a steep density slope \citep{1999ApJ...510..379M,2001ApJ...551..946T}. 
However, for a CSM with a shallow density slope, e.g., a steady wind with $\rho\propto r^{-2}$, shocks usually decelerate as they sweep the surrounding medium \citep{1982ApJ...258..790C,1982ApJ...259..302C}. 
The energy loaded in the shocked gas while the shock is propagating in the interior of the CSM can also contribute to emission in the early phase of llGRBs
Recently, we have developed a semi-analytic model for relativistic ejecta-CSM interaction \citep{2017ApJ...834...32S}, which can be used for estimating the amount of the energy produced by the hydrodynamic interaction.

Recently, the new GRB 171205A was detected by the {\it Swift} satellite \citep{2017GCN.22177....1D}. 
The reported $T_{90}$ duration and the $15$--$50$ keV fluence of the burst were $T_{90}=189.4$ s and $3.6\pm0.3\times10^{-6}$ erg cm$^{-2}$. 
The optical afterglow was immediately identified, and found to be associated with a spiral galaxy at $z=0.0368$ \citep{2017GCN.22180....1I}. 
Assuming the distance of $167$ Mpc, the isotropic equivalent energy of the prompt gamma-ray emission is $1.2\times 10^{49}$ erg, which is much smaller than those of standard GRBs\footnote{The recent paper by \cite{delia2018}, which is published while this manuscript was being peer-reviewed, performed a combined analysis of the {\it Swift} and Konus-Wind datasets and reported a slightly higher isotropic gamma-ray energy, $E_\mathrm{iso}=2.18^{+0.63}_{-0.50}\times 10^{49}$ erg, and a longer $T_{90}=190.5\pm33.9$ s. 
However, the dataset used in this paper is not significantly different from theirs, leaving the conclusions unchanged. }.

From these gamma-ray properties, GRB 171205A is unambiguously classified as an llGRB.  
Observations in other wavelengths have also been carried out by several groups. 
Follow-up optical photometric and spectroscopic observations identified an SN component in the optical afterglow at only $2.4$ days after the discovery \citep{2017GCN.22204....1D}. 
The early spectra of the SN component showed SN 1998bw-like spectral features\footnote{See, also, the recent paper by \cite{2018arXiv181003250W}}. 
Radio observations were initiated about $20$ hours after the trigger, revealing a bright radio source \citep{2017GCN.22187....1D}. 
The prompt {\it Swift} detection and the dedicated multi-wavelength follow-up campaigns have made GRB 171205A one of the most densely observed nearby GRB-SNe.

Development of theoretical light curve models self-consistently explaining the multi-wavelength light curves of GRB 171205A would greatly help us constraining the progenitor scenario for llGRBs and the mechanism responsible for the high-energy emission. 
So far, \cite{2017arXiv171209319D} have attempted theoretical interpretation of the prompt and afterglow light curves of GRB 171205A in the framework of an off-axis SN-GRB. 
In this work, we perform light curve modeling of GRB 171205A based on the relativistic ejecta-CSM interaction model developed by our previous work \citep[two separated papers: ][hereafter \hyperlink{sms17}{SMS17} and \hyperlink{sm18}{SM18}]{2017ApJ...834...32S,2018MNRAS.478..110S}. 
We found that the relativistic ejecta-CSM interaction model can successfully explain the multi-wavelength light curves of GRB 171205A. 
From the light curve modeling, we require some conditions on the dynamical properties of the ejecta produced by the stellar explosion associated with GRB 171205A and the density structure of the CSM surrounding the progenitor star. 
Then, we further discuss the population of llGRBs and other X-ray transients in the framework of the CSM interaction scenario. 

This paper is organized as follows. 
In Section \ref{sec:emission_model}, we describe our emission model. 
Results of the light curve modeling are presented in Section \ref{sec:results}. 
We discuss the implications of the results and the origin of llGRBs in Section \ref{sec:discussion}. 
Finally, Section \ref{sec:conclusion} concludes this paper. 

\section{Emission model}\label{sec:emission_model}
\begin{figure}
\begin{center}
\includegraphics[scale=0.41,bb=0 0 591 481]{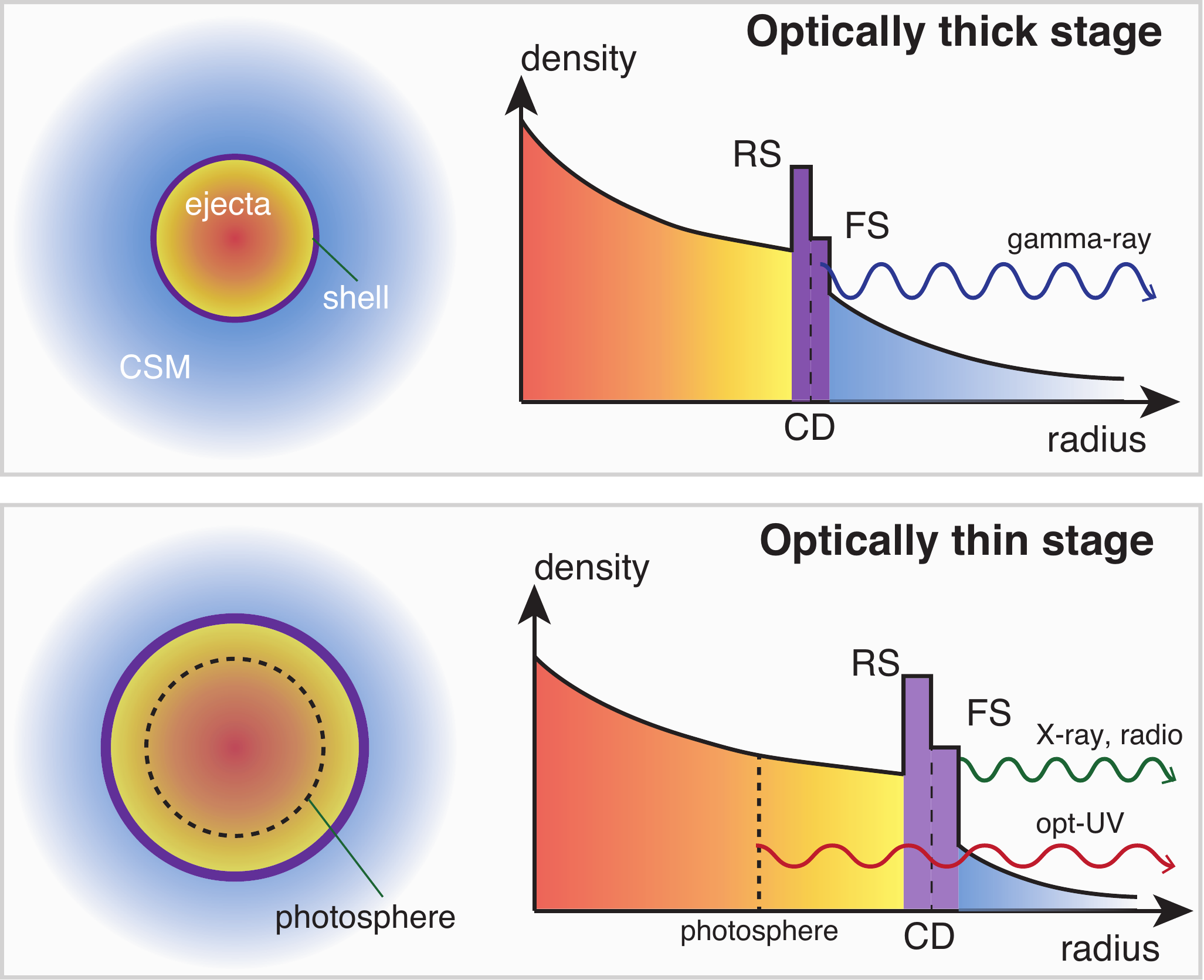}
\caption{Schematic view of the emission model. 
The upper and lower panels correspond to the stages where the shocked gas is optically thick and thin. }
\label{fig:schematic}
\end{center}
\end{figure}

The emission model is based on our previous work (\hyperlink{sms17}{SMS17} and \hyperlink{sm18}{SM18}) with some updates. 
Figure \ref{fig:schematic} schematically represents the situation considered here. 
A massive star explodes in the surrounding CSM and creates expanding spherical ejecta. 
As a result of the collision between the ejecta and the CSM, the gas swept up by the forward and reverse shock fronts forms a geometrically thin shell. 
At early stages of the dynamical evolution, the shell is optically thick (upper panel). 
Thus, the radiation produced by the shock dissipation is basically trapped in the shell and gradually escapes via radiative diffusion as the optical depth of the shell drops down to unity, which we observe as the prompt gamma-ray and X-ray emission. 
The shell becomes transparent at later epochs (lower panel)
Then, thermal photons leaking through the photosphere, which is now receding into the un-shocked ejecta, are observed as optical and UV emission. 
In addition, non-thermal electrons produced in the shock fronts give rise to radio and X-ray emission by synchrotron and inverse Compton processes. 
In the following, we describe our emission model used for the light curve modeling.

\subsection{Hydrodynamics}\label{sec:hydrodynamics}
In \hyperlink{sms17}{SMS17}, we present semi-analytic formulae for the dynamical evolution of mildly relativistic ejecta interacting with a CSM with spherical symmetry. 
The density structure of the ejecta is of fundamental importance for the subsequent dynamical evolution. 
The following several parameters characterize the density structures. 
Since the ejecta are freely expanding, the radial velocity $c\beta$, where $c$ is the speed of light, is given by the radial coordinate $r$ divided by the elapsed time $t$,
\begin{equation}
\beta=\frac{r}{ct}.
\end{equation}
We assume that the radial density profile of the ejecta is given by a broken power-law function of the 4-velocity, $\Gamma\beta$, where $\Gamma=(1-\beta^2)^{-1/2}$ is the Lorentz factor corresponding to the radial velocity. 
The power-law exponent of the outer density profile is denoted by $n$, while the inner density structure is assumed to be flat. 
Thus, the density profile is expressed as follows,
\begin{equation}
\rho_\mathrm{ej}(r,t)=
\left\{\begin{array}{l}
\rho_0\left(\frac{t}{t_0}\right)^{-3}(\Gamma_\mathrm{br}\beta_\mathrm{br})^{-n}\\
\hspace{2em}\mathrm{for}\ \ \ \Gamma\beta\leq \Gamma_\mathrm{br}\beta_\mathrm{br}\\
\rho_0\left(\frac{t}{t_0}\right)^{-3}(\Gamma\beta)^{-n}\\
\hspace{2em}\mathrm{for}\ \ \ \Gamma_\mathrm{br}\beta_\mathrm{br}<\Gamma\beta\leq\Gamma_\mathrm{max}\beta_\mathrm{max},
\end{array}\right.
\label{eq:density_profile}
\end{equation}
where $t_0$ is the initial time of the interaction and we set $t_0=1$ s and $\Gamma_\mathrm{br}\beta_\mathrm{br}=0.1$. 
The maximum Lorentz factor is set to be $\Gamma_\mathrm{max}=5$ throughout this work. 
Note that results are not sensitively dependent on the value of the maximum Lorentz factor, because layers around the maximum velocity are immediately swept by the reverse shock. 
The normalization constant $\rho_0$, the exponent $n$, and the break velocity $\beta_\mathrm{br}$ are important parameters specifying how much energy is loaded in the outermost ejecta. 
These parameters are deeply related to how the relativistic ejecta are produced and connected to non-relativistic supernova ejecta at the bottom, thereby being highly uncertain. 
In the following we assume a relatively flat density profile ($n=5$ for the fiducial model), which realizes a large energy dissipation rate at early epochs, in order to explain early bright emission. 
The normalization constant $\rho_0$ is determined by specifying the kinetic energy of the relativistic part $(\Gamma\beta\geq1)$ of the ejecta,
\begin{equation}
E_\mathrm{rel}=4\pi c^5t^3\int_{1/\sqrt{2}}^{\beta_\mathrm{max}}\rho_\mathrm{ej}(r,t)\Gamma(\Gamma-1)\beta^2d\beta,
\end{equation}
where the lower bound, $\beta=1/\sqrt{2}$, is the velocity corresponding to $\Gamma\beta=1$. 
We use the normalized kinetic energy $E_\mathrm{rel,51}=E_\mathrm{rel}/(10^{51}\ \mathrm{erg})$ as a free parameter to specify the normalization constant $\rho_0$. 
In a similar way, the mass of the relativistic ejecta is defined as follows,
\begin{equation}
M_\mathrm{rel}=4\pi (ct)^3\int_{1/\sqrt{2}}^{\beta_\mathrm{max}}\rho_\mathrm{ej}(r,t)\Gamma\beta^2d\beta.
\end{equation}

At the initial time of the interaction $t=t_0$, the outermost layer of the ejecta is adjacent to the CSM at $r=c\beta_\mathrm{max}t_0$. 
We assume a wind-like CSM with a constant mass-loss rate and a wind velocity,
\begin{equation}
\rho_\mathrm{csm}(r)=Ar^{-2}. 
\end{equation}
We adopt the following normalization for the CSM density parameter $A$, $A_\star=A/(5\times 10^{11}\ \mathrm{g}\ \mathrm{cm})$, and treat the non-dimensional quantity $A_\star$ as a free parameter characterizing the CSM density. 
For a typical wind velocity of compact stars ($v_\mathrm{w}\simeq 10^3$ km s$^{-1}$), $A_\star=1$ corresponds to a mass-loss rate of $\dot{M}\simeq 10^{-5}\ M_\odot$ yr$^{-1}$, which is a typical value for a galactic Wolf-Rayet star. 
For late-time light curve modeling, we also explore the possibility that a dense CSM is present only in the immediate vicinity of the progenitor star, while the outer CSM density is similar to a wind with $\dot{M}=10^{-5}\ M_\odot$ yr$^{-1}$. 
In such cases, we assume that the dense CSM only extends up to $r=r_\mathrm{out}$, 
\begin{equation}
\rho_\mathrm{csm}(r)=\left\{
\begin{array}{ccl}
Ar^{-2}
&r\leq r_\mathrm{out},\\
A_\mathrm{out}r^{-2}
&r_\mathrm{out}<r,
\end{array}\right.
\label{eq:csm_double_power}
\end{equation}
with $A_\mathrm{out}<A$. 
The outer CSM density is also normalized so that $A_{\mathrm{out},\star}=A_\mathrm{out}/(5\times 10^{11}\ \mathrm{g}\ \mathrm{cm})$. 

When the ejecta expand into the CSM, the hydrodynamic interaction between the two media creates forward and reverse shocks propagating in the CSM and the ejecta, converting the kinetic energy of outer layers of the ejecta into the internal energy of the shocked gas. 
The basic idea adopted in this model is the so-called ``thin shell approximation'', in which we treat the region between the two shock fronts as a geometrically thin shell, as in the non-relativistic treatment \citep{1982ApJ...258..790C,1982ApJ...259..302C}. 
We obtain the shell radius $R_\mathrm{s}$ by solving the equation of motion for the thin shell. 
Accordingly, the evolutions of other hydrodynamic quantities, such as the forward and reverse shock radii $R_\mathrm{fs}$ and $R_\mathrm{rs}$, the corresponding shock velocities $c\beta_\mathrm{fs}$ and $c\beta_\mathrm{rs}$, the shell mass $M_\mathrm{s}$, the shell velocity and the Lorentz factor $c\beta_\mathrm{s}$ and $\Gamma_\mathrm{s}$, and the post-shock pressures $P_\mathrm{fs}$ and $P_\mathrm{rs}$ at the forward and reverse shock fronts, are obtained by using the shock jump conditions. 
We have confirmed that the temporal evolution of these hydrodynamic quantities are in good agreement with hydrodynamic simulations for various sets of the free parameters (\hyperlink{sms17}{SMS17}). 

\begin{figure*}
\begin{center}
\includegraphics[scale=0.6,bb=0 0 680 509]{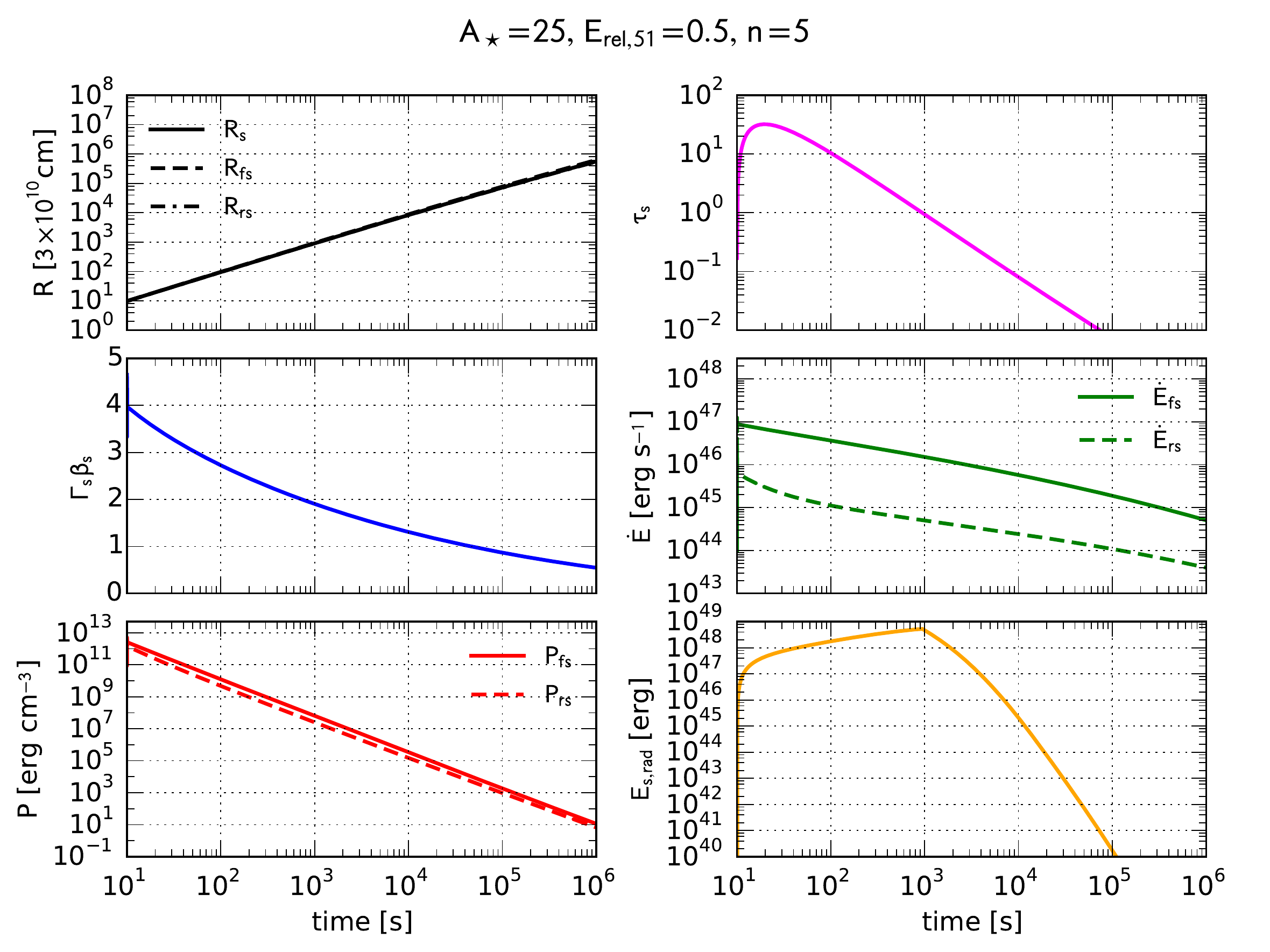}
\caption{Dynamical evolution of the shell caused by the ejecta-CSM interaction. 
In the left panels, we plot the temporal evolutions of the shell and the shock radii ($R_\mathrm{s}$, $R_\mathrm{fs}$, and $R_\mathrm{rs}$), the 4-velocity of the shell ($\Gamma_\mathrm{s}\beta_\mathrm{s}$), and the post-shock pressures ($P_\mathrm{fs}$ and $P_\mathrm{rs}$) from top to bottom. 
The right panels show those of the optical depth of the shell ($\tau_\mathrm{s}$), the energy dissipation rates at the shocks ($\dot{E}_\mathrm{fs}$ and $\dot{E}_\mathrm{rs}$), and the radiation energy of the shell ($E_\mathrm{s,rad}$). 
The parameters of the ejecta and the CSM are set to $E_\mathrm{rel,51}=0.5$, $n=5$, and $A_\star=25$. 
}
\label{fig:dynamics}
\end{center}
\end{figure*}

Figure \ref{fig:dynamics} shows an example of the semi-analytic calculations. 
The free parameter specifying the hydrodynamic model are set to $E_\mathrm{rel,51}=0.5$, $n=5$, and $A_\star=25$. 
The shell and shock radii monotonically increase with time. 
As seen in the top left panel of Figure \ref{fig:dynamics}, their temporal evolutions are almost identical with each other. 
This means that the width between the forward and reverse shock fronts is much smaller than the shell radius, justifying the thin shell approximation. 
As the shell sweeps the CSM, the mass loading and the pressure gradient force within the shell decelerate the shell, resulting in the decreasing shell velocity. 
The post-shock pressures at the shock fronts also decrease with time as the velocity and the pre-shock density decrease.

\subsection{Emission from optically thick shell}
The shocked gas in the thin shell between the forward and reverse shock fronts is optically thick in the early stage of the dynamical evolution. 
We define the optical depth of the shell as follows,
\begin{equation}
\tau_\mathrm{s}=\frac{\kappa M_\mathrm{s}}{4\pi R_\mathrm{s}^2},
\label{eq:tau_sh}
\end{equation}
where $\kappa$ is the opacity and set to $\kappa=0.2$ cm$^2$ g$^{-1}$. 
During the optically thick stage ($\tau_\mathrm{s}>1$), photons in the shell gradually escape from the shell via radiative diffusion. 
In our emission model, photons diffusing out from the shell are regarded as the dominant source of the prompt gamma-ray emission. 
We calculate the bolometric light curve of the emission by adopting the diffusion approximation. 

The forward and reverse shocks convert the kinetic energy of the ejecta into the radiation energy in the shell. 
We assume that the internal energy density in the shell is dominated by radiation. 
The temporal evolution of the radiation energy $E_\mathrm{s,rad}$ of the shell is governed by the balance between the production by the shock dissipation and cooling processes,
\begin{equation}
\frac{dE_\mathrm{s,rad}}{dt}=\dot{E}_\mathrm{fs}+\dot{E}_\mathrm{rs}-\frac{E_\mathrm{s,rad}}{3V}\frac{dV}{dt}-\dot{E}_\mathrm{diff},
\end{equation}
as long as the shell is optically thick, $\tau_\mathrm{s}>1$. 
The 1st and 2nd terms of the R.H.S. of this equation represent the energy production through the forward and reverse shock fronts. 
The energy production rates are given by
\begin{equation}
\dot{E}_\mathrm{fs}=4\pi cR_\mathrm{fs}^2\frac{\gamma_\mathrm{ad}P_\mathrm{fs}}{\gamma_\mathrm{ad}-1}\Gamma_\mathrm{s}^2(\beta_\mathrm{fs}-\beta_\mathrm{s}),
\end{equation}
for the forward shock and
\begin{equation}
\dot{E}_\mathrm{rs}=4\pi cR_\mathrm{rs}^2\frac{\gamma_\mathrm{ad}P_\mathrm{rs}}{\gamma_\mathrm{ad}-1}\Gamma_\mathrm{s}^2(\beta_\mathrm{s}-\beta_\mathrm{rs}),
\end{equation}
for the reverse shock (\hyperlink{sms17}{SMS17}), where $\gamma_\mathrm{ad}=4/3$ is the adiabatic exponent. 
These terms are set to $\dot{E}_\mathrm{fs}=\dot{E}_\mathrm{rs}=0$ after the shell becomes optically thin, $\tau_\mathrm{s}<1$ (instead, the dissipated energy partly contributes to non-thermal emission). 
The 3rd term represents adiabatic cooling, which is proportional to the fractional change in the shell volume $V$ per unit time. 
Finally, the 4th term represents the energy loss due to the radiative diffusion and is given by the following formula,
\begin{equation}
\dot{E}_\mathrm{diff}
=4\pi R_\mathrm{s}^2u_\mathrm{s,rad}v_\mathrm{diff},
\label{eq:dEdt_rad}
\end{equation}
where $u_\mathrm{s,rad}=E_\mathrm{s,rad}/V$  is the radiation energy density of the shell. 
The diffusion velocity $v_\mathrm{diff}$ is obtained from the radiative transfer equation in the diffusion limit,
\begin{equation}
v_\mathrm{diff}=
\frac{c(1-\beta_\mathrm{s}^2)}{(3+\beta_\mathrm{s}^2)\tau_\mathrm{s}+2\beta_\mathrm{s}},
\label{eq:v_diff}
\end{equation}
(see, Appendix of \hyperlink{sms17}{SMS17} for detail). 
The diffusion velocity should not exceed the following maximum value,
\begin{equation}
v_\mathrm{diff,max}=c(1-\beta_\mathrm{s}),
\end{equation}
above which the radiation energy goes through the shell in a superluminal way. 
Therefore, when the velocity calculated by Equation (\ref{eq:v_diff}) is larger than this threshold, we set $v_\mathrm{diff}=v_\mathrm{diff,max}$. 
In other words, this prescription corresponds to the flux-limited diffusion in radiation hydrodynamics, although we deal with only a single zone. 

The temporal evolutions of the optical depth $\tau_\mathrm{s}$, the energy production rates at the forward and reverse shock fronts, and the radiation energy of the shell are presented in the right panels of Figure \ref{fig:dynamics}. 
Initially, the optical depth steeply rises owing to the rapid accumulation of mass. 
After reaching the maximum value, it steadily decreases down to $\tau_\mathrm{s}<1$ as the shell radius increases. 
The shell becomes optically thin at $t\simeq 10^3$ s for this parameter set. 
The energy production rate due to the shock dissipation continuously decreases both for the forward and reverse shocks. 
The energy dissipated at the forward shock front is larger than that at the reverse shock front by more than one order of magnitude, indicating that the non-thermal emission from the forward shock dominates over the reverse shock counterpart. 
Because of the less significant contribution from the reverse shock, we only calculate the non-thermal emission from the forward shock in the following light curve modeling. 
The radiation energy of the shell continues to increase with time as long as the shell is optically thick. 
After the shell becomes optically thin, the shock dissipation no longer contributes to the increase in the radiation energy. 
Then, it decreases owing to the radiative diffusion and adiabatic cooling.

We use the radiative loss rate $\dot{E}_\mathrm{diff}$ to calculate the bolometric luminosity of the emission seen by a distant observer. 
We assume that the intensity of the emission is isotropic in the rest frame of the shell and use the method described in Appendix \ref{sec:light_curve}, which takes into account relativistic effects, to obtain the bolometric luminosity seen in the observer frame. 
The observed bolometric luminosity of the diffusive emission is obtained as follows:
\begin{equation}
L_\mathrm{diff}(t_\mathrm{obs})=
c\int\frac{\dot{E}_\mathrm{diff}(t)}{R_\mathrm{s}(t)\Gamma_\mathrm{s}^3[1-\mu\beta_\mathrm{s}(t)]^3}dt,
\end{equation}
where $\mu$ is given by Equation (\ref{eq:mu}). 

\subsection{Photospheric emission}\label{sec:photospheric_emission}
We also consider the photospheric emission from the pre-shocked SN ejecta. 
After the shell becomes optically thin, the photosphere recedes into deeper layers of the SN ejecta (Figure \ref{fig:schematic}). 
Inner layers in the pre-shocked ejecta successively become transparent and release the remaining internal energy as thermal photons. 
We adopt the following simplified model to calculate the luminosity and the temperature of the photospheric emission. 

First, the optical depth between the forward shock front and a layer with a velocity $r/t$ in the pre-shocked SN ejecta at $t$ is calculated as follows,
\begin{equation}
\tau(t,r)=\tau_\mathrm{s}+\int_{r}^{R_\mathrm{rs}(t)}\kappa\rho_\mathrm{ej}(r,t)dr. 
\end{equation}
The 1st term in the R.H.S. is the contribution from the shell, Equation (\ref{eq:tau_sh}), which is now less than unity. 
The 2nd term is the contribution from the pre-shocked SN ejecta, which are now truncated by the reverse shock at $r=R_\mathrm{rs}$. 
The photospheric radius $R_\mathrm{ph}(t)$ at $t$ is determined so that $\tau(t,R_\mathrm{ph})=1$ is satisfied. 

The ejecta are supposed to originate from a stellar atmosphere swept by a blast wave generated as a result of the core collapse. 
The shock kinetic energy is converted to the kinetic energy and internal energy of the downstream gas, which leads to a comparable fraction of the kinetic and thermal energy contents within the ejecta before significant adiabatic loss. 
We assume that the initial internal energy distribution, $u_\mathrm{ej}(t_0,r)$, of the SN ejecta is proportional to the kinetic energy distribution, $\Gamma(\Gamma-1)\rho_\mathrm{ej}(t_0,r)$. 
We introduce a constant $f_\mathrm{th}$ specifying the ratio of the internal energy density to the kinetic energy density at $t=t_0(=1\ \mathrm{s})$, i.e., $u_\mathrm{ej}(t_0,r)=f_\mathrm{th}\Gamma(\Gamma-1)\rho_\mathrm{ej}(t_0,r)$. 
Each layer of the ejecta loses the internal energy by adiabatic expansion.  
Specifically, the internal energy density should decrease as $u_\mathrm{ej}(t,0)\propto t^{-3\gamma_\mathrm{ad}}$. 
Thus, the internal energy density profile $u_\mathrm{ej}(t,r)$ is described as follows,
\begin{equation}
u_\mathrm{ej}(t,r)=\left(\frac{t}{t_0}\right)^{-3\gamma_\mathrm{ad}}f_\mathrm{th}\Gamma(\Gamma-1)\rho_\mathrm{ej}(t_0,r)c^2.
\end{equation}
In the following, we assume a fixed value of $f_\mathrm{th}=0.3$ so that the observed optical and UV fluxes can be explained. 
The amount of the radiation energy released within a small time interval from $t$ to $t+\Delta t$ is calculated as follows. 
The photospheric radius evolves from $R_\mathrm{ph}(t)$ to $R_\mathrm{ph}(t+\Delta t)\simeq R_\mathrm{ph}(t)+dR_\mathrm{ph}/dt\Delta t$ within the time interval. 
The layer corresponding to the photosphere at $t$ travels at the velocity $R_\mathrm{ph}(t)/t$ and then reaches $r=R_\mathrm{ph}(t)(t+\Delta t)/t$ at $t+\Delta t$. 
Therefore, the volume $\Delta V$ that newly becomes transparent is given by
\begin{eqnarray}
\Delta V
&=&\frac{4\pi}{3}\left[R_\mathrm{ph}(t)^3\left(\frac{t+\Delta t}{t}\right)^3-R_\mathrm{ph}(t+\Delta t)^3\right]
\nonumber\\
&\simeq& 4\pi R_\mathrm{ph}(t)^2\left[\frac{R_\mathrm{ph}(t)}{t}-\frac{dR_\mathrm{ph}}{dt}\right]\Delta t.
\end{eqnarray}
The internal energy lost through the photosphere per unit time yields
\begin{equation}
\dot{E}_\mathrm{ph}(t)
=4\pi R_\mathrm{ph}(t)^2
\left[\frac{R_\mathrm{ph}(t)}{t}-\frac{dR_\mathrm{ph}}{dt}\right]
u_\mathrm{ej}(t,R_\mathrm{ph}).
\end{equation}
 The observed bolometric luminosity is calculated in the same way as the optically thick shell:
\begin{equation}
L_\mathrm{ph}(t_\mathrm{obs})=
c\int\frac{\dot{E}_\mathrm{ph}(t,\nu)}{R_\mathrm{s}(t)\Gamma_\mathrm{s}^3[1-\mu\beta_\mathrm{s}(t)]^3}dt.
\end{equation}

 In addition, we assume that the radiation is well represented by blackbody emission. 
 The radiation temperature at the photosphere is obtained from the Stefan-Boltzmann law,
 \begin{equation}
 T_\mathrm{ph}=\left[\frac{\dot{E}_\mathrm{ph}(t)}{4\pi R_\mathrm{ph}(t)^2\sigma_\mathrm{SB}}\right]^{1/4},
 \end{equation}
 where $\sigma_\mathrm{SB}$ is the Stefan-Boltzmann constant. 
 Using the radiation temperature as the color temperature of the blackbody emission, we obtain the luminosity per unit frequency,
\begin{equation}
\left(\frac{d\dot{E}}{d\nu}\right)_{\mathrm{ph}}=\frac{2\pi\dot{E}_\mathrm{ph}(t)}{c^2\sigma_\mathrm{SB} T_\mathrm{ph}^4}\frac{h\nu^3}{\exp(h\nu/k_\mathrm{B}T_\mathrm{ph})-1},
\end{equation}
which is used to calculate the observed luminosity per unit frequency,
\begin{equation}
L_{\nu,\mathrm{ph}}(t_\mathrm{obs})=
c\int\frac{(d\dot{E}/d\bar{\nu})_\mathrm{ph}(\bar{\nu})}{R_\mathrm{s}(t)\Gamma_\mathrm{s}^2[1-\mu\beta_\mathrm{s}(t)]^2}dt,
\end{equation}
where $\bar{\nu}$ is the comoving frequency given by Equation (\ref{eq:nu_bar}). 

\subsection{Non-thermal emission}
After the shell becomes optically thin and most of photons trapped in the shell have been released, the shocked gas starts producing non-thermal photons via synchrotron and inverse Compton processes. 
Photospheric photons considered in the previous subsection serve as seed photons for the inverse Compton emission. 
We calculate the non-thermal emission following the early high-energy emission by using the method developed by \hyperlink{sm18}{SM18}. 
We focus on the non-thermal emission from the forward shock, because the energy dissipation rate at the forward shock front dominates over that of the reverse shock (Section \ref{sec:hydrodynamics}).

\subsubsection{Electron momentum distribution}
We treat non-thermal electrons produced at the shock front in one-zone approximation. 
In other words, we assume that these electrons are uniformly distributed in a narrow region close to their production site and do not treat their spatial advection and diffusion. 
Furthermore, we assume that their angular distribution in the momentum space is isotropic. 
Thus, their momentum distribution is expressed as a function of time $t$ and the norm of the momentum $p_\mathrm{e}$, $dN/dp_\mathrm{e}(t,p_\mathrm{e})$. 

The temporal evolution of the electron momentum distribution is obtained by solving the following advection equation in the momentum space for the range from $p_\mathrm{min}=10^{-3}m_\mathrm{e}c$ to $p_\mathrm{max}=10^6m_\mathrm{e}c$,
\begin{equation}
\frac{\partial }{\partial t}\left(\frac{dN}{dp_\mathrm{e}}\right)=
\frac{\partial}{\partial p_\mathrm{e}}\left[(\dot{p}_\mathrm{syn}+\dot{p}_\mathrm{ic}+\dot{p}_\mathrm{ad})\frac{dN}{dp_\mathrm{e}}\right]
+\left(\frac{d\dot{N}}{dp_\mathrm{e}}\right)_\mathrm{in}.
\label{eq:govern}
\end{equation}
We assume that all the electrons swept by the shock front experience non-thermal acceleration process and then obey a power-law momentum distribution with an exponent $-p$, 
\begin{equation}
\left(\frac{d\dot{N}}{dp_\mathrm{e}}\right)_\mathrm{in}\propto 
\left\{
\begin{array}{cl}
p_\mathrm{e}^{-p}&\mathrm{for}\ p_\mathrm{in}\leq p_\mathrm{e}\leq p_\mathrm{max},\\
0&\mathrm{otherwise}.
\end{array}
\right.
\end{equation}
The normalization and the minimum injection momentum $p_\mathrm{in}$ are determined by the energy dissipation rate at the forward shock front and the average electron energy. 
As usually assumed in many non-thermal emission models for GRBs and SNe \citep[e.g.,][]{1998ApJ...497L..17S,2001ApJ...548..787S,2002ApJ...568..820G}, we introduce a parameter $\epsilon_\mathrm{e}$ and assume that a fraction $\epsilon_\mathrm{e}$ of the internal energy of the gas in the downstream of the shock is converted to the energy of non-thermal electrons, $u_\mathrm{ele}=\epsilon_\mathrm{e}u_\mathrm{int}$, where $u_\mathrm{int}$ is the internal energy density at the shock front. 
The average energy of a single non-thermal electron is given by the electron internal energy $u_\mathrm{ele}$ divided by the electron number density $n_\mathrm{ele}$ in the downstream, $u_\mathrm{ele}/n_\mathrm{ele}$. 

The momentum loss rates, $\dot{p}_\mathrm{syn}$ and $\dot{p}_\mathrm{ic}$, due to synchrotron emission and inverse Compton cooling can be calculated from the corresponding energy loss rates. 
They are given by
\begin{equation}
\dot{p}_\mathrm{syn}=\frac{4\sigma_\mathrm{T}u_\mathrm{B}}{3m_\mathrm{e}^2c^2}p_\mathrm{e}\sqrt{m_\mathrm{e}^2c^2+p_\mathrm{e}^2},
\label{eq:pdot_syn}
\end{equation}
and
\begin{equation}
\dot{p}_\mathrm{ic}=\frac{4\sigma_\mathrm{T}u_\mathrm{rad}}{3m_\mathrm{e}^2c^2}p_\mathrm{e}\sqrt{m_\mathrm{e}^2c^2+p_\mathrm{e}^2},
\label{eq:pdot_ic}
\end{equation}
where $\sigma_\mathrm{T}$ is the Thomson cross section, $u_\mathrm{B}$ the magnetic energy density, and $u_\mathrm{rad}$ the energy density of seed photons. 
We further adopt a frequently-adopted prescription that the magnetic energy density is given by $u_\mathrm{B}=\epsilon_\mathrm{B}u_\mathrm{int}$, where $\epsilon_\mathrm{B}$ is the other microphysics parameter specifying the fraction of the magnetic energy density to the internal energy density. 
As we will see below, the photospheric and synchrotron emission contribute to seed photons for inverse Compton emission. 
Thus, we use the radiation energy densities of photospheric and synchrotron photons for $u_\mathrm{rad}$. 
The adiabatic momentum loss rate is
\begin{equation}
\dot{p}_\mathrm{ad}=\frac{p_\mathrm{e}}{3V}\frac{dV}{dt}.
\label{eq:pdot_ad}
\end{equation}

The advection equation, Equation (\ref{eq:govern}), is numerically solved by a 1st-order implicit upwind scheme.

\subsubsection{Synchrotron emission}
For a given electron momentum distribution, calculations of synchrotron emissivity $j_{\nu,\mathrm{syn}}$ and the self-absorption coefficient $\alpha_{\nu,\mathrm{syn}}$ are straightforward. 
We use the widely used formulae in the literature \citep[e.g.][]{1979rpa..book.....R}:
\begin{equation}
j_{\nu,\mathrm{syn}}=\frac{1}{4\pi V}\int P_{\nu,\mathrm{syn}}(\gamma_\mathrm{e})\frac{dN}{dp_\mathrm{e}}dp_\mathrm{e},
\end{equation}
for the synchrotron emissivity, and
\begin{equation}
\alpha_{\nu,\mathrm{syn}}=\frac{c^2}{8\pi V\nu^2}\int 
\frac{\partial}{\partial p_\mathrm{e}}\left[p_\mathrm{e}\gamma_\mathrm{e}P_{\nu,\mathrm{syn}}(\gamma_\mathrm{e})\right]
\frac{1}{p_\mathrm{e}^2}\frac{dN}{dp_\mathrm{e}}dp_\mathrm{e},
\end{equation}
for the absorption coefficient, where $P_{\nu,\mathrm{syn}}(\gamma_\mathrm{e})$ is the synchrotron power per unit frequency for a single electron with a Lorentz factor of $\gamma_\mathrm{e}=[1+p_\mathrm{e}^2/(m_\mathrm{e}^2c^2)]^{1/2}$. 
The corresponding synchrotron self-absorption optical depth is the product of the absorption coefficient and the shell width $V/(4\pi R_\mathrm{s}^2)$,
\begin{equation}
\tau_\mathrm{\nu,\mathrm{ssa}}=\frac{c^2}{32\pi^2 R_\mathrm{s}^2\nu^2}\int 
\frac{\partial}{\partial p_\mathrm{e}}\left[p_\mathrm{e}\gamma_\mathrm{e}P_{\nu,\mathrm{syn}}(\gamma_\mathrm{e})\right]
\frac{1}{p_\mathrm{e}^2}\frac{dN}{dp_\mathrm{e}}dp_\mathrm{e}. 
\end{equation}
Since the intensity of the synchrotron emission is expressed in the following way,
\begin{equation}
I_\mathrm{syn}(\nu)=\frac{j_{\nu,\mathrm{syn}}}{\alpha_{\nu,\mathrm{syn}}}(1-e^{-\tau_{\nu,\mathrm{ssa}}}),
\end{equation}
the corresponding synchrotron energy loss rate per unit frequency yields
\begin{equation}
\left(\frac{d\dot{E}}{d\nu}\right)_\mathrm{syn}=16\pi^2 R_\mathrm{fs}^2I_\mathrm{syn}(\nu).
\end{equation}
The observed luminosity per unit frequency is given by
\begin{equation}
L_{\nu,\mathrm{syn}}(t_\mathrm{obs})=
2c\int\frac{(d\dot{E}(t,\bar{\nu})/d\bar{\nu})_\mathrm{syn}}{R_\mathrm{s}(t)\Gamma_\mathrm{s}^2[1-\mu\beta_\mathrm{s}(t)]^2}dt.
\end{equation}

\subsubsection{Inverse Compton emission}
The inverse Compton emission is calculated by the following formula,
\begin{equation}
I_\mathrm{ic}(\nu)=\int G(\gamma_\mathrm{e},\nu_\mathrm{i},\nu)\frac{dN}{dp_\mathrm{e}}I_\mathrm{seed}(\nu_\mathrm{i})dp_\mathrm{e}d\nu_\mathrm{i},
\end{equation}
for a given electron momentum distribution $dN/dp_\mathrm{e}$ and a seed photon intensity $I_\mathrm{seed}(\nu_\mathrm{i})$. 
The redistribution function $G(\gamma_\mathrm{e},\nu_\mathrm{i},\nu)$ gives the energy spectrum of scattered photons for incoming mono-energetic electrons with the Lorentz factor $\gamma_\mathrm{e}$ and monochromatic photons with the frequency $\nu_\mathrm{i}$ (see, Appendix of \hyperlink{sm18}{SM18}). 
We also note that the adopted redistribution function correctly takes into account the Klein-Nishina suppression. 
We consider the photospheric emission and the synchrotron emission as the sources of seed photons. 
\begin{equation}
I_\mathrm{seed}(\nu)=\frac{1}{16\pi^2 R_\mathrm{fs}^2}\frac{1-\beta_\mathrm{s}(t)}{1-\beta_\mathrm{ph}(t')}\left(\frac{d\dot{E}}{d\nu}\right)_\mathrm{ph}
+I_\mathrm{syn}(\nu). 
\end{equation}
We note that the contribution from the photospheric emission needs some corrections because of the difference between the photospheric and forward shock radii. 
The time $t'$ appearing in the 1st term of the R.H.S is defined so that a photospheric photon emitted at this time into the radial direction reaches the forward shock at $t$, $R_\mathrm{fs}(t)=R_\mathrm{ph}(t')+c(t-t')$. 
The energy loss rate per unit frequency $(d\dot{E}/d\nu)_\mathrm{ph}$ should be evaluated at this time. 
The term $(1-\beta_\mathrm{s})/(1-\beta_\mathrm{ph})$ is the correction factor for the energy density, where $\beta_\mathrm{ph}(t')=R_\mathrm{ph}(t')/(ct')$ is the velocity of the layer at the photospheric radius at the time $t'$. 

In a similar way to the synchrotron emission, the observed luminosity per unit frequency is given by
\begin{equation}
L_{\nu,\mathrm{ic}}(t_\mathrm{obs})=
2c\int\frac{(d\dot{E}(t,\bar{\nu})/d\bar{\nu})_\mathrm{ic}}{R_\mathrm{s}(t)\Gamma_\mathrm{s}^2[1-\mu\beta_\mathrm{s}(t)]^2}dt,
\end{equation}
with
\begin{equation}
\left(\frac{d\dot{E}}{d\nu}\right)_\mathrm{ic}=16\pi^2 R_\mathrm{fs}^2I_\mathrm{ic}(\nu).
\end{equation}

\section{Results}\label{sec:results}
\subsection{Prompt gamma-ray emission}

\begin{figure}
\begin{center}
\includegraphics[scale=0.4,bb=0 0 576 720]{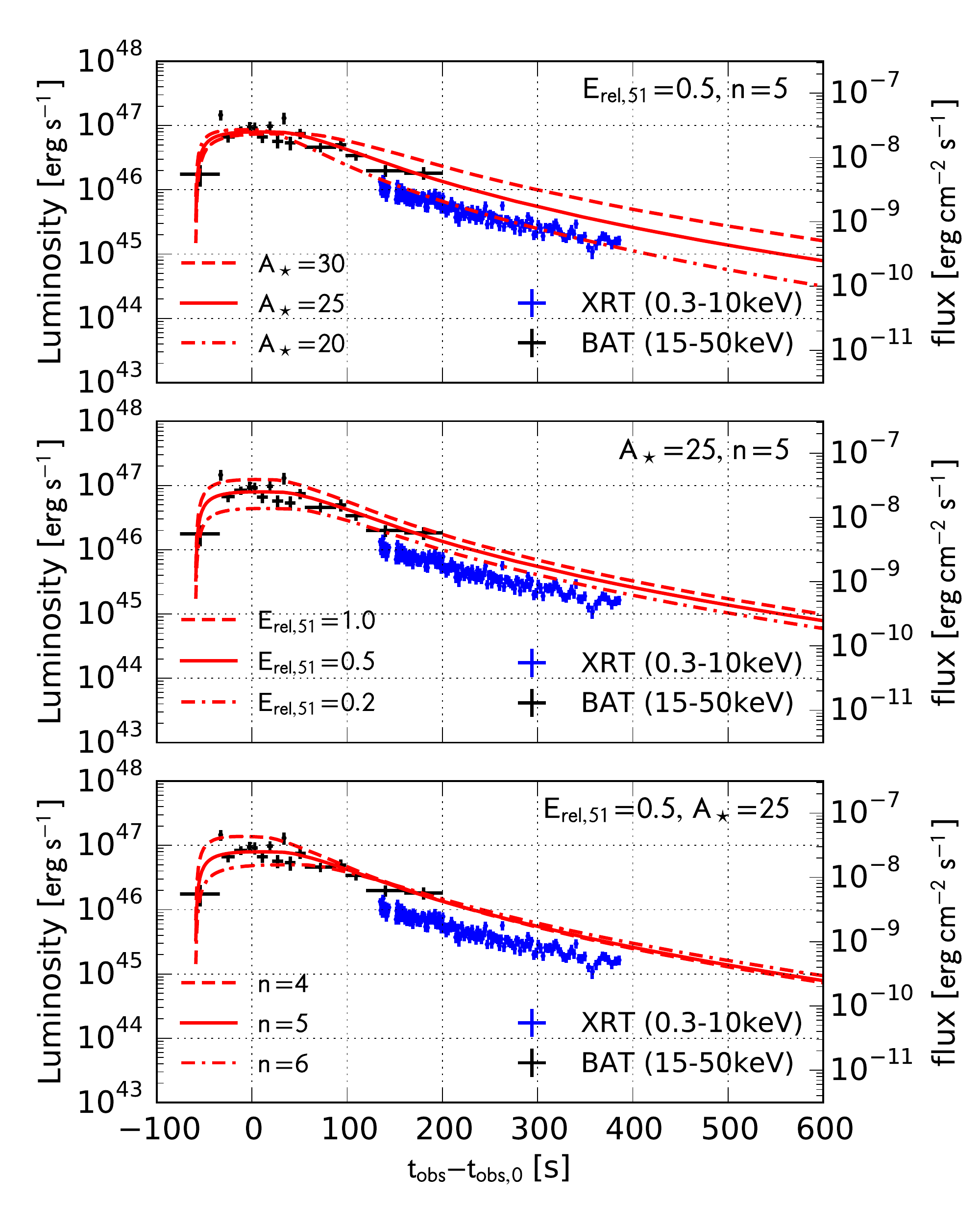}
\caption{Theoretical light curves compared with early gamma-ray and X-ray observations by the {\it Swift}.
The luminosity from $t=-100$ s to $t=+600$ s is shown ($t=0$ corresponds to the BAT trigger).
The black and blue crosses represent the {\it Swift}/BAT ($15$--$50$ keV) and XRT ($0.3$--$10$ keV) observations. 
The theoretical models with different sets of parameters are plotted. 
In all the panels, the fiducial model with $ A_\star=25$, $E_\mathrm{rel,51}=0.5$, and $n=5$ is plotted as a solid line. 
The dashed and dash-dotted lines in each panel represent models with $A_\star=30$ and $20$ (top), $E_\mathrm{rel,51}=1.0$ and $0.2$ (middle), and $n=4$ and $6$ (bottom). }
\label{fig:bat}
\end{center}
\end{figure}

First, we focus on the prompt gamma-ray emission and the subsequent X-ray emission. 
In Figure \ref{fig:bat}, we plot the gamma-ray and X-ray luminosities of GRB 171205A observed by the {\it Swift}/BAT and XRT. 
We make use of the data processed and provided by UK Swift Science Data Centre\footnote{http://www.swift.ac.uk/} \citep[see,][]{2007A&A...469..379E,2009MNRAS.397.1177E}. 
The gamma-ray luminosity first reaches $\sim 10^{47}$ erg s$^{-1}$ and then declines down to $\sim 10^{46}$ erg s$^{-1}$ in $\sim 200$ s. 
The XRT observation has been conducted from $t_\mathrm{obs}\simeq 135$ s. 
The BAT spectrum (from $t_\mathrm{obs}\simeq -40$ s to $200$ s) is well fitted by a single power-law distribution with a photon index of $1.41\pm0.14$ \citep{2017GCN.22184....1B}. 
A slightly softer photon index ($1.717^{+0.035}_{-0.024}$) is reported for the XRT spectrum at later epochs (from $t_\mathrm{obs}=135$ s to $400$ s; \citealt{2017GCN.22183....1K}). 

Theoretical bolometric light curves are compared with the observed light curves in Figure \ref{fig:bat}. 
Our fiducial model assumes $A_\star=25$, $E_\mathrm{rel,51}=0.5$, and $n=5$, which is shown as a solid line in each panel of Figure \ref{fig:bat}. 
Models with different values of $A_\star$, $E_\mathrm{rel,51}$, and $n$ are also plotted in each panel. 
The theoretical light curves show a remarkable agreement with the BAT and XRT light curves. 
We note that the theoretical light curve in specific energy ranges can be different from those shown in Figure \ref{fig:bat}, since the theoretical model cannot produce frequency dependent light curves. 
In particular, the $0.3$--$10$ keV flux from $t_\mathrm{obs}=100$ s to $200$ s, during which both BAT and XRT observations are available, is smaller than the $15$--$ 50$ keV flux by a factor of a few. 
Thus, the bolometric flux must be larger than $0.3$--$10$ keV flux owing to the contribution from photons with higher energies, although the flux of the high-energy photons at later times appears to be below the BAT detection threshold. 
This is also supported by the XRT photon index harder than $2$. 
It is therefore natural that our fiducial bolometric light curve adjusting the BAT result is slightly more luminous than the XRT light curve at $t_\mathrm{obs}>100$ s but shows a similar decay rate. 

\subsection{$E_\mathrm{rad}$--$T_\mathrm{burst}$ diagram}
\begin{figure}
\begin{center}
\includegraphics[scale=0.55,bb=0 0 453 509]{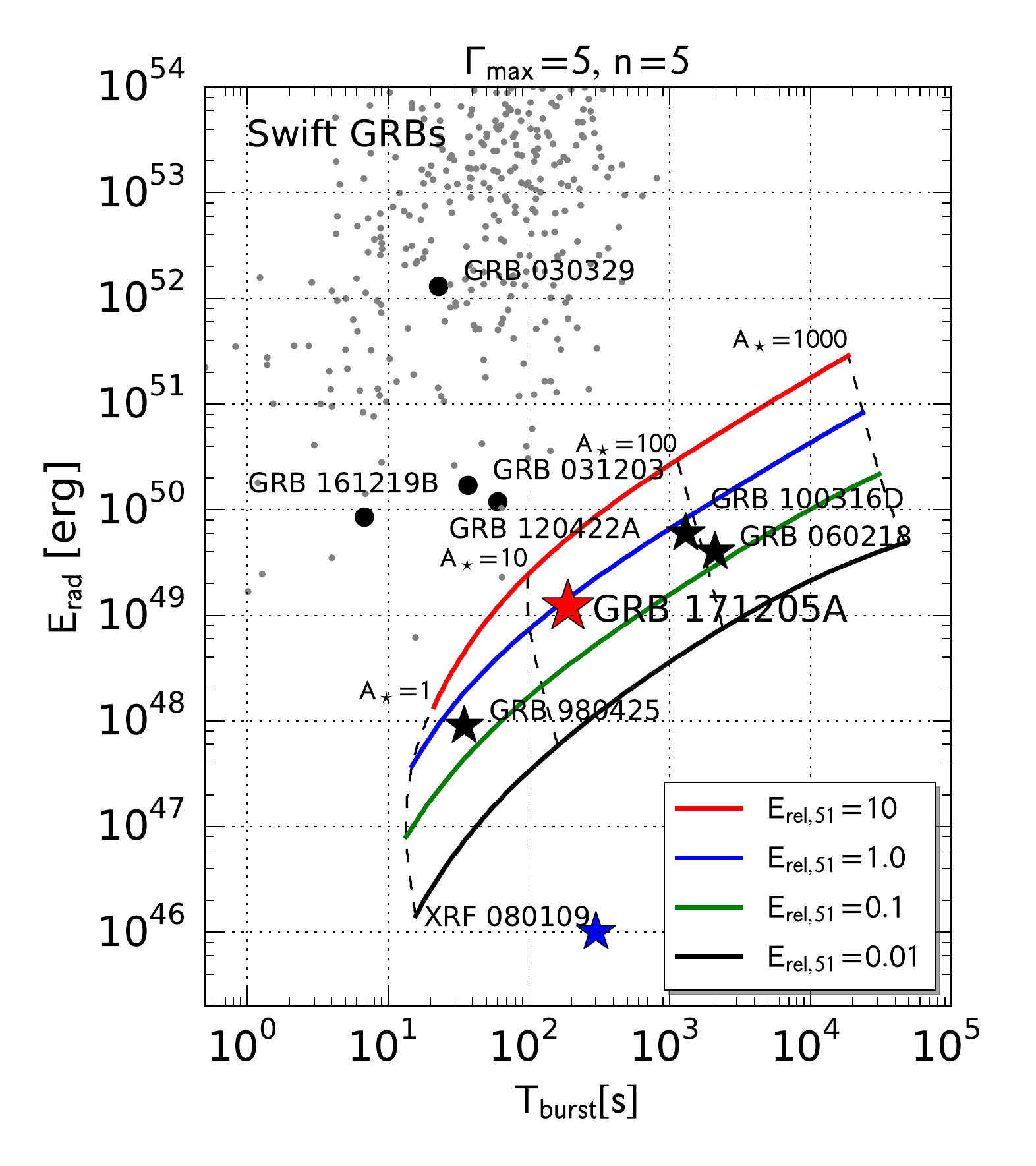}
\caption{Scatter plot showing (isotropically equivalent) radiated energy vs burst duration. 
Stars represent nearby GRBs unambiguously classified as llGRBs, while gray circles represent nearby GRBs with SNe whose gamma-ray properties cannot be explained by the CSM interaction model. 
GRB 171205A is shown by the red star. 
Gray dots correspond to GRBs with spectroscopically confirmed redshifts. 
The data are taken from 3rd {\it Swift}/BAT GRB catalog (https://swift.gsfc.nasa.gov/results/batgrbcat/) compiled by \protect\cite{2016ApJ...829....7L}. 
The solid curves show the relation between the radiated energy and the duration predicted by the CSM interaction model. 
In each curve, the assumed value of the CSM density $A_\star$ increases from $A_\star=1$ to $A_\star=1000$, with $E_\mathrm{rel}$ and $n$ fixed. 
The red, blue, green, and black solid curves (from top to bottom) corresponds to models with $E_\mathrm{rel,51}=10$, $1.0$, $0.1$, $0.01$.  
The adopted values of $A_\star$ are shown as dashed black curves with labels $A_\star=1$, $10$, $100$, and $1000$. }
\label{fig:diagram}
\end{center}
\end{figure}

In Figure \ref{fig:diagram}, we present a $E_\mathrm{rad}$--$T_\mathrm{burst}$ diagram, in which GRBs with known $E_\mathrm{iso}$ and $T_{90}$ are compared with theoretical predictions. 
We plot swift GRBs with known redshifts from 3rd {\it Swift}/BAT GRB catalog compiled by \cite{2016ApJ...829....7L}. 
They occupy the upper left region of the diagram, reflecting their large $E_\mathrm{iso}$ and short $T_{90}$. 
On the other hand, llGRBs, GRB 980425, 060218, 100316D, and 171205A, are located in the lower right region. 
Some nearby GRBs associated with SNe, GRB 030329, 031203, 120422A, and 161219B, are also plotted. 

In Figure \ref{fig:diagram}, we plot theoretical predictions as done by \hyperlink{sms17}{SMS17}. 
The radiated energy $E_\mathrm{rad}$ is calculated by integrating the theoretical bolometric light curve with respect to time. 
The burst duration $T_\mathrm{burst}$ is defined as the observer time at which $90\%$ of the total radiated energy has been received. 
The solid curves show the relations between the radiated energy $E_\mathrm{rad}$ and the duration $T_\mathrm{burst}$ predicted by theoretical models with different kinetic energies of the ejecta. 
Each curve is obtained by increasing the CSM density parameter $A_\star$ from $A_\star=1$ to $A_\star=1000$. 
As discussed by \hyperlink{sms17}{SMS17}, locations of llGRBs are successfully explained by the ejecta-CSM interaction scenario, while some GRBs with SNe (GRB 030329, 031203, 120422A, and 161218B) are not (this will be discussed in Section \ref{sec:llGRBs}). 
With the new llGRB 171205A included, llGRBs seem to follow the trend that llGRBs with longer durations produce larger amounts of isotropic gamma-ray energy. 
This trend is also in line with the theoretical expectation that larger amounts of CSM lead to more prolonged emission with larger radiated energies. 
Swift GRBs are well above the theoretical curves, clearly indicating that a highly collimated emission region, i.e., an ultra-relativistic jet, is required to explain their large isotropic equivalent gamma-ray energies released over short durations.

\subsection{Photospheric emission}
\begin{figure*}
\begin{center}
\includegraphics[scale=0.6,bb=0 0 648 360]{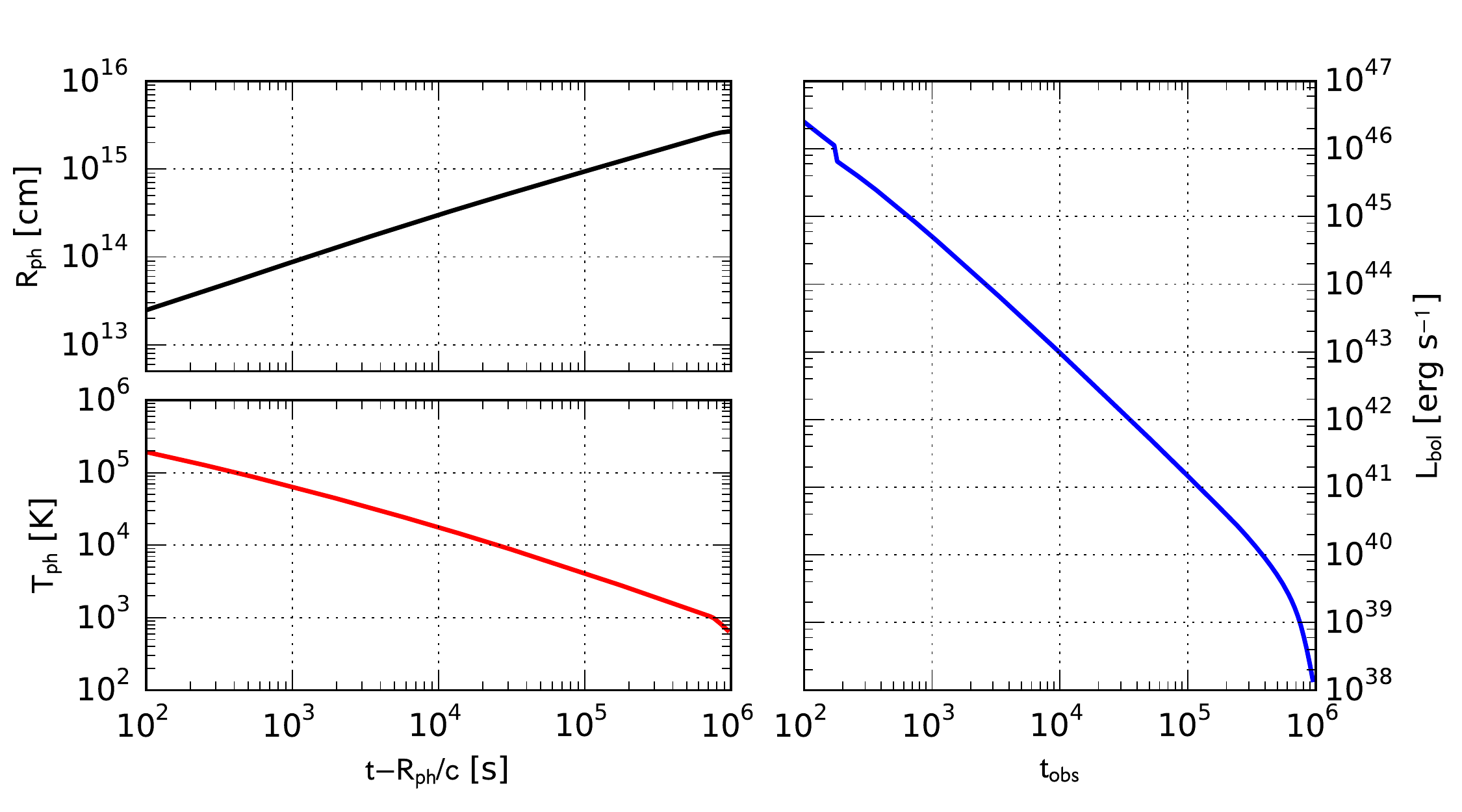}
\caption{Properties of the photospheric emission. 
In the left column, the photospheric radius $R_\mathrm{ph}$ (top) and temperature $T_\mathrm{ph}$ (bottom) are plotted as a function of the delay time $t-R_\mathrm{ph}(t)/c$. 
In the right column, we plot the bolometric luminosity as a function of the observer time. 
}
\label{fig:thermal}
\end{center}
\end{figure*}

The photospheric emission expected after the optically thick stage predominantly contributes to optical-UV emission observed by the UVOT telescope on board the {\it Swift} satellite. 
Figure \ref{fig:thermal} presents the temporal behaviors of the photospheric emission. 
The left panels of Figure \ref{fig:thermal} show the photospheric radius and the temperature as a function of the delay time $t-R_\mathrm{ph}(t)/c$, which roughly corresponds to the observer time. 
On the other hand, the right panel of Figure \ref{fig:thermal} represents the bolometric light curve, which is given as a function of the observer time $t_\mathrm{obs}$. 
The photospheric radius steadily increases as the ejecta expand. 
Optically thick layers stratified below the photosphere cool via adiabatic expansion, which results in the monotonically declining bolometric luminosity and the photospheric temperature.  

\begin{figure}
\begin{center}
\includegraphics[scale=0.6,bb=0 0 360 576]{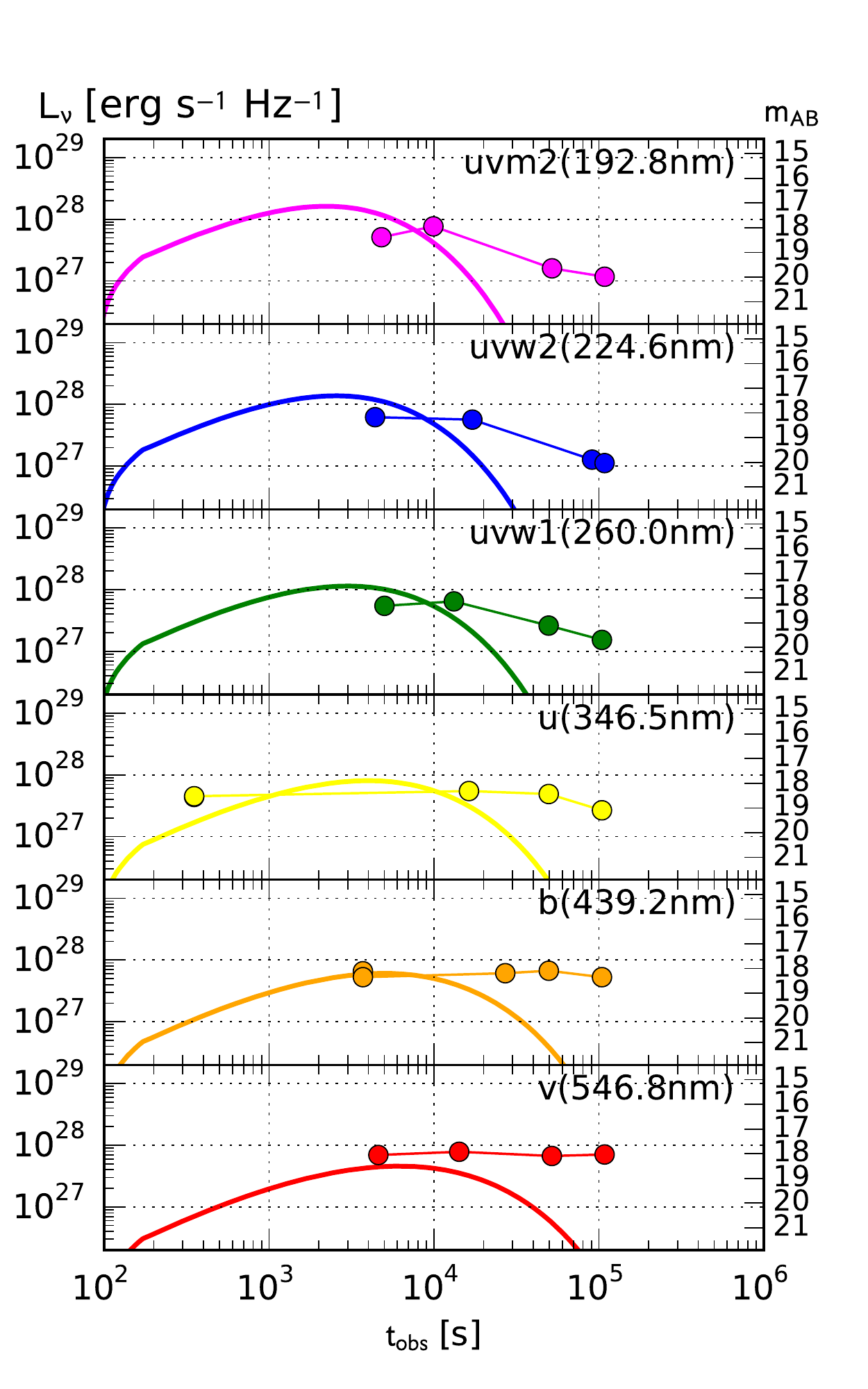}
\caption{Multi-band light curve of the photospheric emission. 
In each panel, we plot the luminosity per unit frequency $L_{\nu,\mathrm{ph}}$ for the frequencies corresponding to the central wavelengths of the UVOT filters ({\it uvm2}, {\it uvw2}, {\it uvw1}, {\it u}, {\it b}, and {\it v}-bands from top to bottom). 
The theoretical multi-band light curves (solid lines) are compared with the UVOT observations (filled circles; \citealt{2017GCN.22202....1S}). 
}
\label{fig:uvot}
\end{center}
\end{figure}

In Figure \ref{fig:uvot}, we plot the multi-band light curves of the photospheric emission. 
The theoretical luminosity per unit frequency is calculated for the wavelengths of $\lambda=190$, $220$, $260$, $350$, $440$, and $550$ nm, which roughly correspond to the central wavelengths of the UVOT filters. 
The multi-color light curve is compared with the UVOT observations up to $10^5$ s. 
The AB magnitudes in the UVOT bands are obtained from the reported magnitudes \citep{2017GCN.22202....1S} and the UVOT zero-points \citep{2008MNRAS.383..627P,2011AIPC.1358..373B}. 
We corrected the magnitudes by assuming $R_V=3.1$ and the Galactic extinction of $E(B-V)_\mathrm{MW}=0.05$ \citep{2017GCN.22202....1S} and using the formula provided by \cite{1989ApJ...345..245C}. 
Although the theoretical multi-band light curve reproduces similar magnitudes to those of the UVOT observations at $t_\mathrm{obs}\simeq10^4$ s, their temporal evolutions do not fully explain the UVOT observations. 
The observed UV fluxes (uvw1, uvw2, and uvm2) increases from $t_\mathrm{obs}\simeq 5\times 10^3$ s to $10^4$ s, while the theoretical light curves decline. 
This discrepancy is probably due to the simplified treatment of the photospheric emission. 
UV emission from SNe is known to suffer from various transfer effects, such as line-blanketing \citep[e.g.,][]{2010ApJ...721.1608B}. 
Moreover, we note that the assumption of the electron scattering opacity for a fully ionized gas is no longer valid at photospheric temperatures of several thousands K or less. 
In addition, the theoretical light curve expects less luminous UV and optical fluxes at $t_\mathrm{obs}\simeq 10^5$ s than the observations. 
This is probably caused by the contribution of other power source(s). 
Several ground-based observations have found the re-brightening of this event mainly in optical and IR bands within 2 days after the trigger \citep{2017GCN.22204....1D}, which is interpreted as the emergence of the associated SN. 
Since we do not include contribution from any other power source, especially radioactive nuclei, the theoretical multi-color light curves continue to declines even after $10^5$ s. 

\subsection{Non-thermal emission}
In the following, we present results of the non-thermal emission modelling. 
The microphysics parameters used in the calculations are set to $\epsilon_\mathrm{e}=0.1$, $\epsilon_\mathrm{B}=0.01$, and $p=3.0$ \citep[e.g.,][]{1999ApJ...526..716L,2015MNRAS.448..417B,2015ApJ...805..164N}. 

\subsubsection{Broad-band light curve}
\begin{figure}
\begin{center}
\includegraphics[scale=0.45,bb=0 0 566 680]{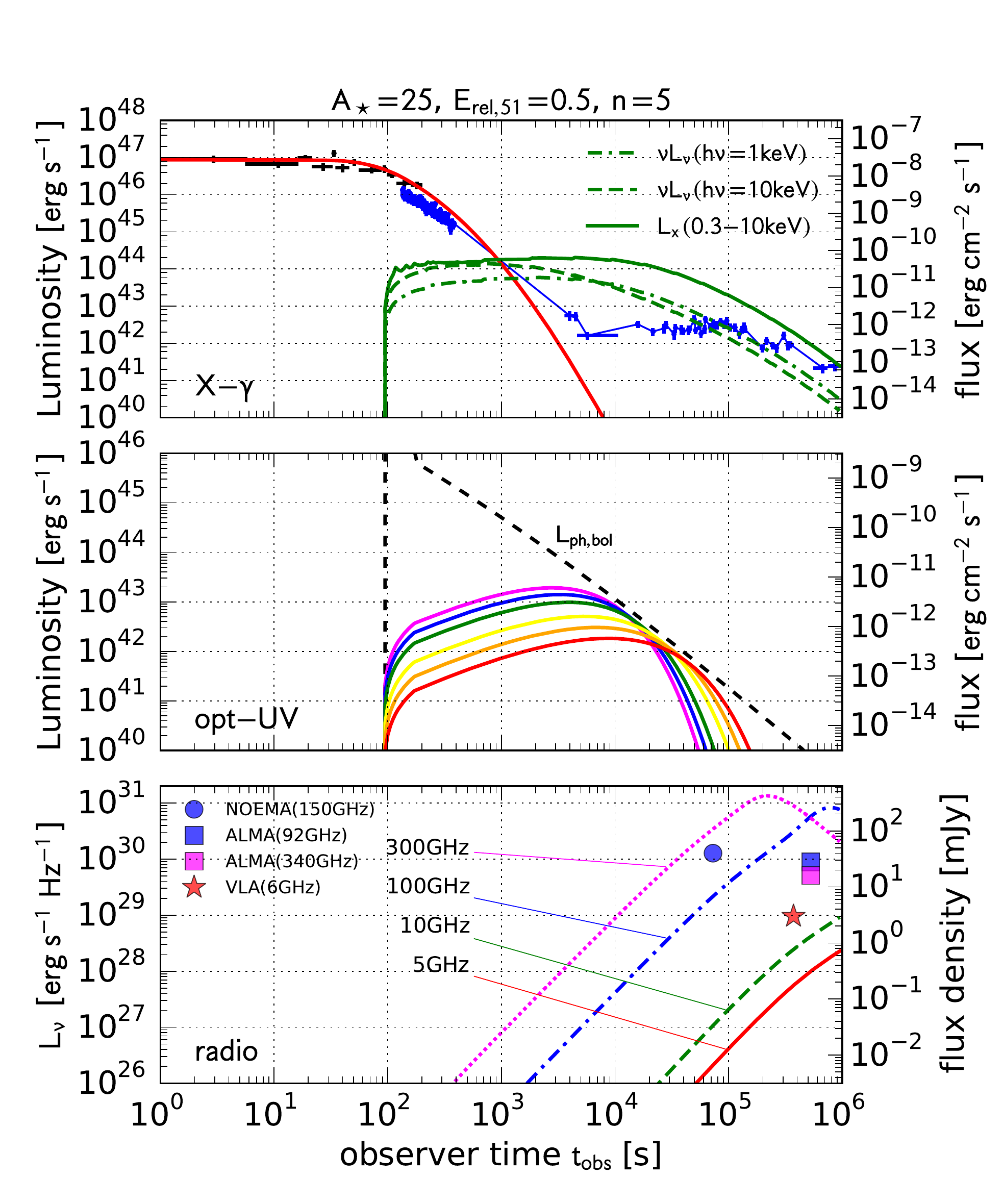}
\caption{X-ray, optical, and radio light curves calculated by our emission model. 
The light curves in X-ray (top panel), optical-UV (middle panel), and radio (bottom panel) bands are compared with observations of GRB 171205A. 
In the top panel, we plot the BAT and XRT observations (the same symbols as Figure \ref{fig:bat}). 
The light curve shown as a red solid is the fiducial model shown in Figure \ref{fig:bat}. The late X-ray emission is dominated by the inverse Compton emission. 
While the green solid line shows the $0.3$--$10$ keV light curve of the inverse Compton emission, the $\nu L_\nu$ light curves at $1$ and $10$ keV are plotted as dashed and dash-dotted lines. 
In the middle panel, we plot the same multi-band light curves as Figure \ref{fig:thermal} in $\nu L_\nu$ as well as the bolometric light curve $L_\mathrm{ph,bol}$ (dashed line). 
In the bottom panel, radio light curves at $5$ (solid), $10$ (dashed), $100$ (dash-dotted), and $300$ (dotted) GHz are compared with early radio observations by NOEMA \protect\citep[blue circle;][]{2017GCN.22187....1D}, ALMA \protect\citep[blue and magenta squares;][]{2017GCN.22252....1P}, and VLA \protect\citep[red star;][]{2017GCN.22216....1L}.
}
\label{fig:broadband_lc}
\end{center}
\end{figure}

In Figure \ref{fig:broadband_lc}, the theoretical light curves in X-ray, optical-UV, and radio bands are shown and compared with multi-wavelength observations of GRB 171205A. 
At first, we assume the same free parameters for the ejecta and the CSM as the fiducial model in the previous section, $A_\star=25$, $E_\mathrm{rel,51}=0.5$, and $n=5$. 
The early emission in the optically thick stage is also plotted in the top panel showing the BAT and XRT light curves. 
The optical and UV light curves shown in the middle panel are the same model as Figure \ref{fig:thermal}, but in $\nu L_\nu$. 

\begin{figure}
\begin{center}
\includegraphics[scale=0.45,bb=0 0 566 680]{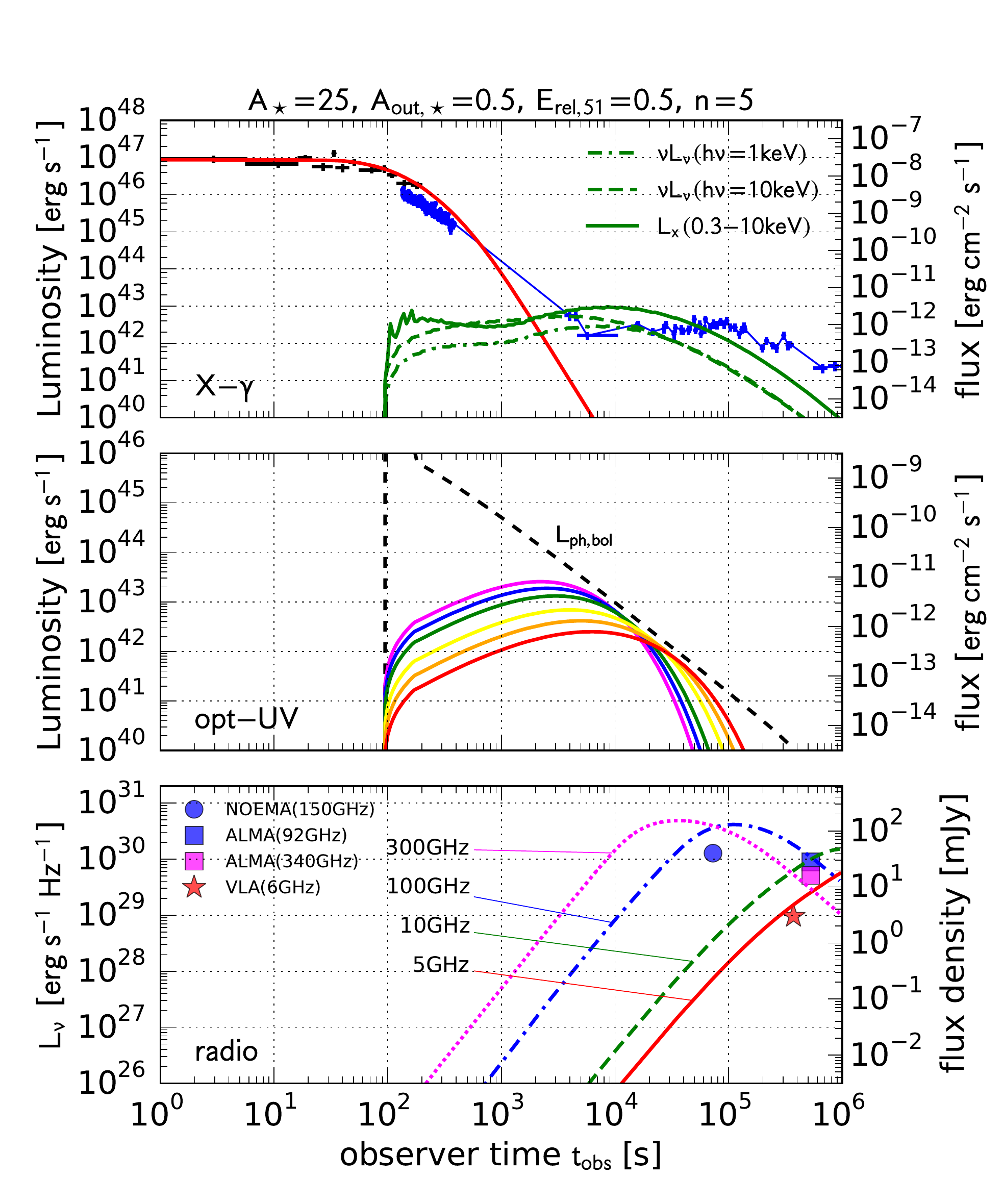}
\caption{Same as Figure \ref{fig:broadband_lc}, but with the reduced outer CSM density of $A_{\mathrm{out},\star}=0.5$. }
\label{fig:broadband_lc2}
\end{center}
\end{figure}

As is clearly seen in the top panel of Figure \ref{fig:broadband_lc}, the theoretical light curve significantly overestimates the X-ray luminosity. 
The X-ray emission is dominated by the inverse Compton emission of non-thermal electrons, whose flux is proportional to the product of the energy density of seed photons and the number density of non-thermal electrons. 
Since the seed photons are predominantly produced by the photospheric emission and its flux is constrained by the UVOT observations, the number density of non-thermal electrons or equivalently the electron energy injection rate at the shock front should be modified for alleviating this disagreement. 
One adjustable parameter is the fraction $\epsilon_\mathrm{e}$. 
However, reducing $\epsilon_\mathrm{e}$ leads to smaller minimum injection momenta $p_\mathrm{in}$. 
For a significantly small $p_\mathrm{in}$, the X-ray break frequency, above which the spectral energy distribution softens, can be in the observed energy range of $0.3$--$10$ keV or even lower, contradicting the observed X-ray spectrum with the hard photon index. 
One possible solution to this discrepancy is relaxing the constraint on the CSM density parameter $A_\star$ obtained by the gamma-ray light curve fitting. 
In this ejecta-CSM interaction model, the prompt gamma-ray emission probes the CSM extending up to $r\sim3\times 10^{13}$ cm from the center. 
While a relatively dense CSM is required in the immediate vicinity of the star, the CSM density beyond this region can be lower than expected by a simple extrapolation of the inverse square law, $\rho_\mathrm{csm}\propto r^{-2}$. 
The implication for this structure will further be discussed in Section \ref{sec:progenitor}. 
Therefore, we explore the possibility that the CSM density drops beyond $r_\mathrm{out}=3\times 10^{13}$ cm, while keeping the inner CSM density fixed. 
In other words, we employ Equation (\ref{eq:csm_double_power}) with reduced outer CSM densities $A_\mathrm{out}<A$. 
In Figure \ref{fig:broadband_lc2}, we plot the multi-wavelength light curves with the outer CSM density of $A_{\mathrm{out},\star}=0.5$, which is smaller by a factor of $50$ than the model shown in Figure \ref{fig:broadband_lc}. 
The early emission from the optically thick shell and the photospheric emission are almost identical with the fiducial model because the parameters of the SN ejecta remain unchanged. 
The late-time X-ray and radio light curves are better reproduced by this model with modified CSM structure.

The theoretical $0.3$--$10$ keV light curve in Figure \ref{fig:broadband_lc2} exhibits a plateau from $t_\mathrm{obs}\simeq 10^3$ s to $2\times 10^{4}$ s with a luminosity of $\simeq 10^{43}$ erg s$^{-1}$. 
Although the plateau X-ray luminosity is still larger than the observed X-ray luminosity by a factor of a few, the plateau and the subsequent decay broadly reproduce the observed features. 
The theoretical X-ray light curve appears to decline faster than the XRT light curve after $t_\mathrm{obs}=10^5$ s. 
This may be improved by including radioactively powered thermal emission, which additionally provides seed photons for the inverse Compton emission. 
The radio light curves in several bands are also plotted in the bottom panel of Figure \ref{fig:broadband_lc2}. 
The flux density in each band initially rises and then declines in a power-law fashion, and its peak appears earlier for higher frequencies. 
These trends are common for young radio-emitting SNe, where the rising and declining parts correspond to optically thick and thin synchrotron emission, respectively \citep[e.g.,][]{2016arXiv161207459C}. 
The $100$ and $350$ GHz radio flux densities reach the peak values of $\sim 100$ mJy at $t_\mathrm{obs}=3\times 10^4$ s and $10^5$ s. 
The fluxes continue to decline with $\sim t^{-1.5}$ after the peaks. 
At $t_\mathrm{obs}\simeq 5\times 10^5$ s, ALMA observations were carried out at $92$ and $340$ GHz. 
The reported flux densities of a few 10 mJy \citep{2017GCN.22252....1P} are roughly consistent with the theoretical fluxes at similar frequencies of $100$ and $300$ GHz. 
Since the flux density at $92$ GHz is smaller than that at $340$ GHz, the synchrotron spectrum in this frequency range is likely to have been in the optically thin regime. 
In this regime, the spectral slope depends on the assumed exponent $p$ of the electron momentum distribution, $\propto \nu^{-p/2}$ or $\nu^{-(p-1)/2}$. 
On the other hand, radio fluxes at lower frequencies, $5$ and $10$ GHz, are still rising even at $t_\mathrm{obs}=10^{6}$ s, which is also consistent with Karl G. Jansky Very Large Array (VLA) observations by \cite{2017GCN.22216....1L}, claiming a spectral slope consistent with a synchrotron self-absorbed spectrum.

\subsubsection{Electron momentum distribution}\label{sec:electron_distribution}
\begin{figure}
\begin{center}
\includegraphics[scale=0.55,bb=0 0 453 339]{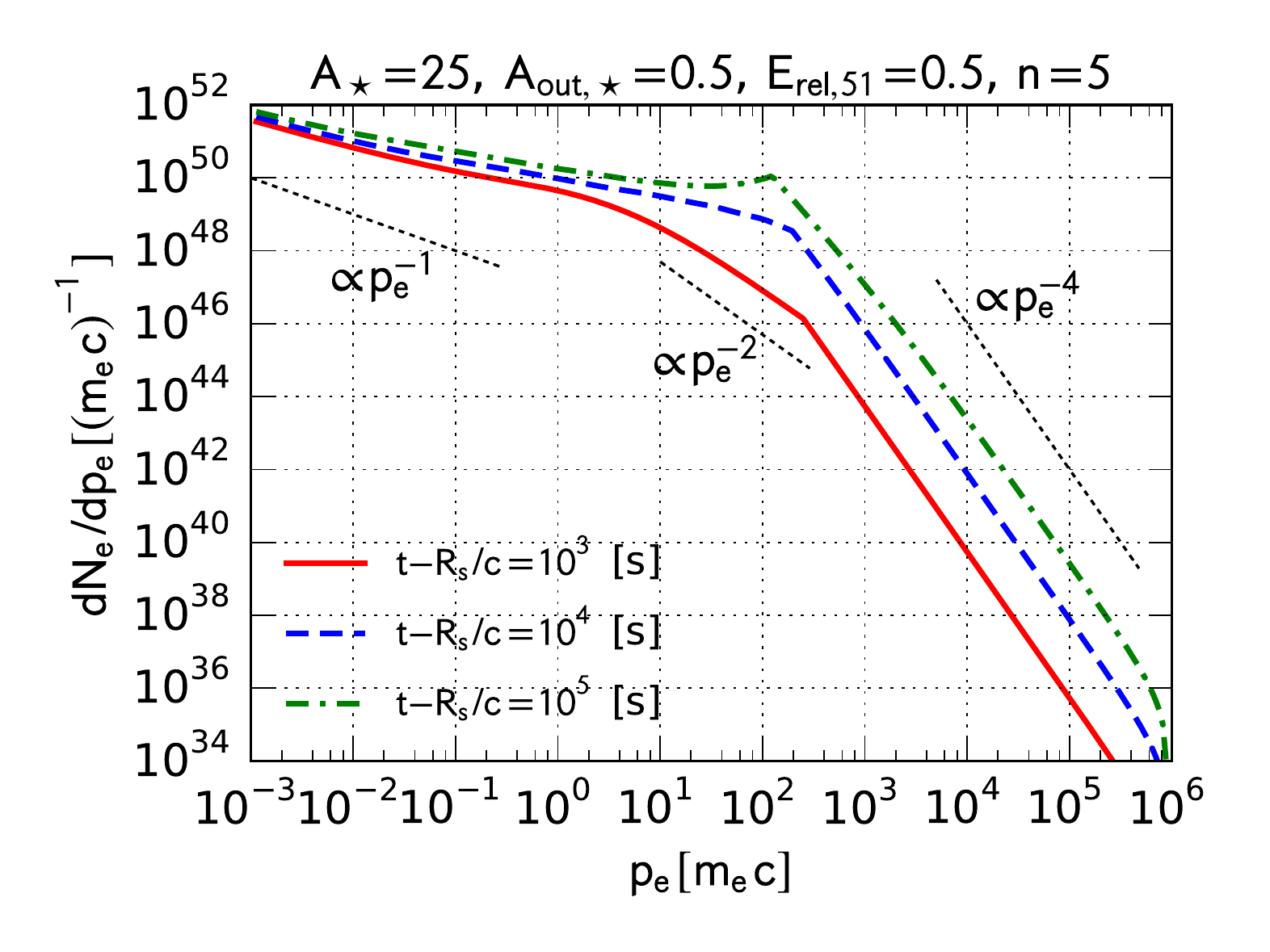}
\caption{Electron momentum distributions at several epochs. 
The solid, dashed, and dash-dotted curves represent the distributions when the elapsed time and the shell radius satisfy $t-R_\mathrm{s}/c=10^{3}$, $10^4$, and $10^5$ s. 
Power-law distributions expected in different regimes are shown in thin dotted lines.}
\label{fig:espec}
\end{center}
\end{figure}

Figure \ref{fig:espec} shows the electron momentum distributions at several epochs. 
The plotted electron distributions are those at epochs satisfying $t-R_\mathrm{s}(t)/c=10^{3}$, $10^4$, and $10^5$ s. 
At these epochs, non-thermal electrons with the plotted momentum distributions predominantly contribute to the non-thermal emission at the observer times of $t_\mathrm{obs}=10^3$, $10^4$, and $10^5$ s. 

The distributions at early epochs are generally a broken power-law function with three segments, the high energy part with the spectral slope of $d\ln N/d\ln p_\mathrm{e}=-4$, the low energy part with a flat slope, and the intermediate part between them. 
The high energy part is composed of electrons with the momentum higher than the minimum injection momentum at several $10m_\mathrm{e}c$. 
These electrons suffer from efficient Compton cooling, which makes the slope steeper by $-1$ than that of the injected electrons, $dN/dp_\mathrm{e}\propto p_\mathrm{e}^{-p-1}$, i.e., the fast cooling regime. 
Electrons having cooled further down to energies below the minimum injection momentum constitute the intermediate part. 
At the earliest epochs of $t-R_\mathrm{s}(t)/c=10^{3}$ s, the spectral slope in this part is $-2$, $dN/dp_\mathrm{e}\propto p_\mathrm{e}^{-2}$, which is also realized in the fast cooling regime for electrons with energies lower than the minimum injection energy. 
As we will see below, this spectral slope is important in determining the photon index of the X-ray spectra. 
At the low energy part ($p_\mathrm{e}<m_\mathrm{e}c$), a flatter spectral slope is realized.  
For $p_\mathrm{e}\ll m_\mathrm{e}c$, all the cooling terms expressed by Equations (\ref{eq:pdot_syn}), (\ref{eq:pdot_ic}), and (\ref{eq:pdot_ad}), are proportional to $p_\mathrm{e}$. 
Therefore, in the steady state and for $p_\mathrm{e}<p_\mathrm{in}$, where the time-dependence and the injection term of Equation (\ref{eq:govern}) should vanish, the resulting governing equation,
\begin{equation}
\frac{\partial}{\partial p_\mathrm{e}}\left[\left( \dot{p}_\mathrm{syn}+\dot{p}_\mathrm{ic}+\dot{p}_\mathrm{ad}\right)\frac{dN}{dp_\mathrm{e}}\right]=0,
\end{equation}
requires $dN/dp_\mathrm{e}\propto p_\mathrm{e}^{-1}$. 
These power-law functions expected in the different regimes of the momentum distribution are also represented in Figure \ref{fig:espec}. 

\subsubsection{Spectral energy distribution}\label{sec:sed}
\begin{figure}
\begin{center}
\includegraphics[scale=0.50,bb=0 0 504 576]{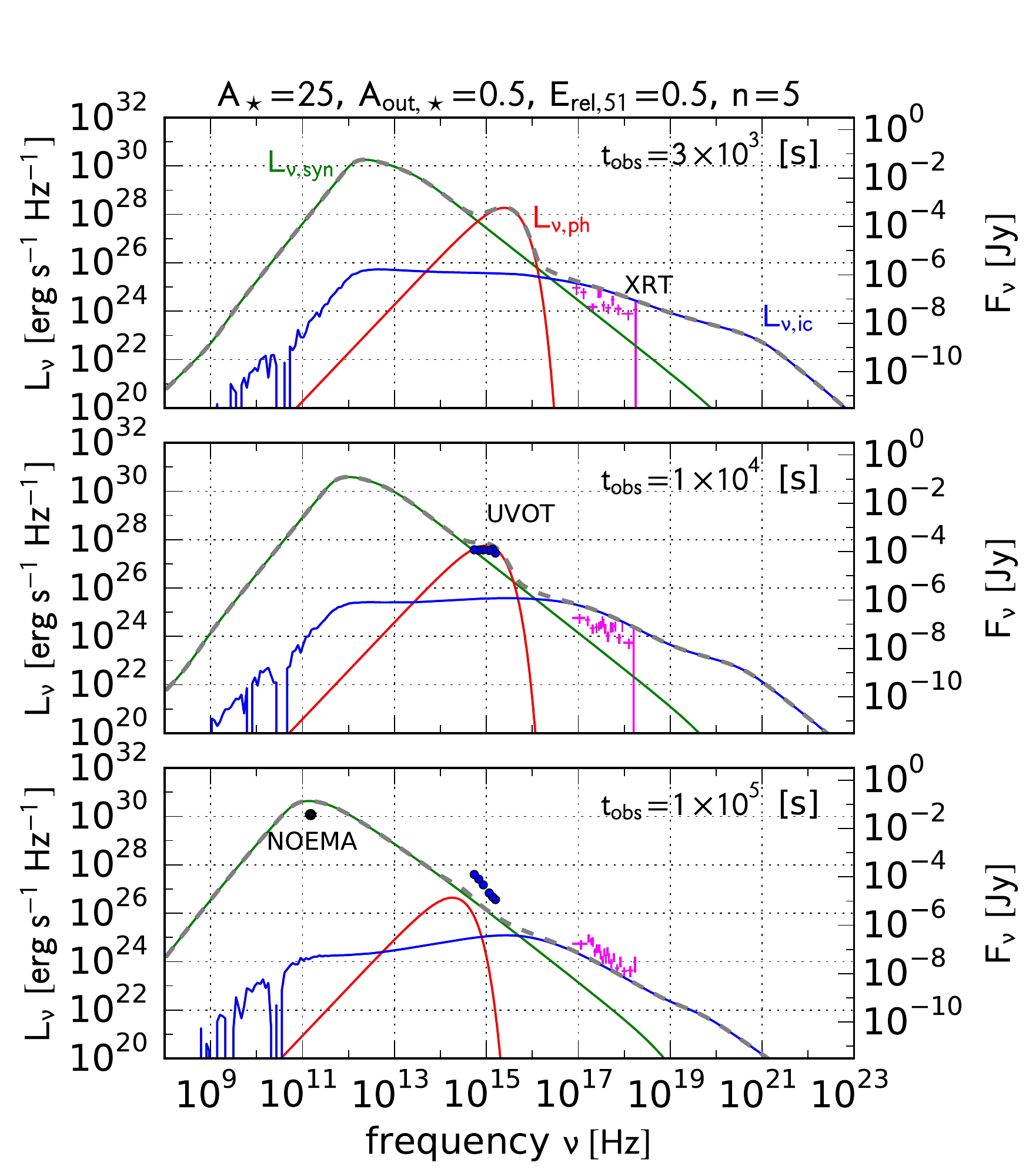}
\caption{Spectral energy distributions at $t_\mathrm{obs}=3\times 10^{3}$, $10^4$, and $10^5$ s from top to bottom. 
The luminosity per unit frequency (or equivalently the flux density) is plotted for the synchrotron (green), photospheric (red), and inverse Compton (blue) components. 
Some observational data are plotted for comparison. 
The black circle in the bottom panel represents NOEMA observation. 
The UVOT multi-color photometric data are plotted as blue circles in the middle and bottom panels. 
The time-sliced XRT data after absorption correction (see Appendix \ref{sec:xrt}) are plotted as magenta crosses in all the panels. }
\label{fig:sed}
\end{center}
\end{figure}

Figure \ref{fig:sed} shows the broad-band spectral energy distributions ($\nu L_\nu$) at $t_\mathrm{obs}=3\times 10^{3}$, $10^4$, and $10^5$ s.  
The contributions of the synchrotron emission, the photospheric emission, and the inverse Compton emission, each of which dominates the total emission at radio, optical, and X-ray (or gamma-ray) energy ranges, are separately shown in each panel. 
First, the synchrotron component dominates the radio flux and is well represented by a broken-power law function, whose spectral slope is $L_\nu\propto \nu^{5/2}$ and $\nu^{-p/2}$ in low and high frequency parts. 
Second, the peak of the photospheric emission appears in the optical-UV range and the spectral energy distribution is given by a Planck function as we have assumed in Section \ref{sec:photospheric_emission}. 
Finally, the inverse Compton emission is the convolution of the electron momentum distribution with the synchrotron and photospheric photon spectra, resulting in relatively complex spectral energy distributions. 
The inverse Compton emission in the X-ray energy range is created by photospheric photons scattered by non-thermal electrons with energies close to the minimum injection energy. 

The spectra in the X-ray range of $0.3$--$10$ keV show power-law distributions with hard photon indices, $\Gamma_\mathrm{ph}\simeq 1.5$. 
The photon index reflects the slope of the electron momentum distribution below the minimum injection energy, i.e., the intermediate part (Section \ref{sec:electron_distribution}). 
Since electrons in this part are predominantly produced by inverse Compton cooling, the electron energy spectrum follows $\gamma_\mathrm{e}^{-2}$, which leads to the inverse Compton spectrum with the photon index of $\Gamma_\mathrm{ph}=1.5$. 
The X-ray spectrum gradually softens with time and the photon index around $0.3$ keV becomes nearly $\Gamma_\mathrm{ph}\simeq2$ at $t_\mathrm{obs}=10^5$ s. 

Some observational data at similar epochs, NOEMA \citep{2017GCN.22187....1D} and UVOT \citep{2017GCN.22202....1S}, are plotted in Figure \ref{fig:sed} and compared with the theoretical spectral energy distributions. 
For the X-ray emission, we have obtained time-sliced X-ray spectra in the $0.3$--$10.0$ keV energy range by analyzing the XRT data (see, Appendix \ref{sec:xrt} for detail). 
The spectral fitting by an absorbed single power-law function have been performed and the results are summarized in Table \ref{table:fitting}. 
The absorption corrected X-ray spectra for the three different time intervals, $t_\mathrm{obs}=10^3$ s to $10^4$ s, $10^4$ s to $3\times 10^4$ s, and $10^5$ s to $2\times 10^{5}$ s, are plotted in Figure \ref{fig:sed}. 
As seen in Figure \ref{fig:sed}, the theoretical spectral energy distributions show overall agreements with current observational constraints in radio, optical, and X-ray. 
The slopes of the absorption corrected X-ray spectra are well explained by the theoretical model. 
The best-fit photon indices of the XRT spectra at earlier two time intervals are $\Gamma_\mathrm{ph}\sim 1.7$ or $1.8$, while it softens to $\Gamma_\mathrm{ph}\simeq 2.3$ at $t_\mathrm{obs}>10^5$ s. 
The spectral softening expected by the theoretical model appears to agree with the temporal evolution of the X-ray spectra, although observations with better photon statistics are needed.

\section{Discussion}\label{sec:discussion}
\subsection{Broad-band emission from llGRBs}
The broad-band light curves calculated by our theoretical model successfully explain several key properties of GRB 171205A. 
The early gamma-ray and X-ray light curves of GRB 171205A are reproduced by the emission diffusing out from the optically thick shell, which is a natural consequence of the relativistic ejecta-CSM interaction. 
The light curve fitting suggests the ejecta kinetic energy of $5\times 10^{50}$ erg and the CSM density parameter of $A_\star=25$. 
After the shell becomes transparent, photospheric and non-thermal emission contribute to the subsequent multi-wavelength emission. 
The photospheric emission from the ejecta expects optical and UV fluxes comparable to observed values. 
We found that a wind-like CSM based on a simple extrapolation of the density profile of $A_\star r^{-2}$ with $A_\star=25$ to outer radii leads to a too bright X-ray afterglow. 
In order to ease the discrepancy, we introduced a sudden change in the CSM density at $r_\mathrm{out}=3\times 10^{13}$ cm and obtained an X-ray light curve marginally consistent with observations.

We note that the discrepancy might also be resolved by a more sophisticated treatment of radiative transfer in the shell. 
For example, the theoretical model does not treat photons re-processed by the CSM after diffusing out from the shell. 
Such photons are expected to be scattered by the CSM and reach the observer with delays \citep[e.g.,][]{2015ApJ...805..159M}. 
In addition, the hydrodynamic model assumes spherical symmetry. 
Although the spherical model succeeds in explaining the properties of the gamma-ray emission, both the ejecta and the CSM can be asymmetric, which could further modify the early gamma-ray and afterglow light curves. 
As expected in the shock emergence in massive stars, asymmetric shock fronts affect the shock breakout light curve in various ways \citep{2010ApJ...717L.154S,2013ApJ...779...60M,2016ApJ...825...92S,2018ApJ...853...52O,2018ApJ...856..146A}. 
Although llGRBs do not require highly collimated ultra-relativistic jet owing to their low gamma-ray luminosities, deviations from spherical symmetry are naturally expected to some extent.

\subsection{Origin of relativistic ejecta and progenitor system}\label{sec:progenitor}
The successful light curve fitting indicates that the stellar explosion responsible for GRB 171205A was associated with the creation of an ejecta component traveling at mildly relativistic speeds. 
The required kinetic energy of the ejecta with 4-velocities faster than the speed of light ($\Gamma\beta\geq 1$), $E_\mathrm{rel}=5\times 10^{50}$ erg, is already comparable to the canonical explosion energy of CCSNe. 
Ordinary CCSNe powered by the neutrino mechanism are unlikely to produce such relativistic ejecta components, clearly suggesting that some additional mechanisms should operate in depositing energy into a small amount of stellar materials. 
For the density profile adopted in our model, Equation (\ref{eq:density_profile}),  with $E_\mathrm{rel}=5\times 10^{50}$ erg, the mass of the relativistic component is only $M_\mathrm{rel}=4.5\times 10^{-5}\ M_\odot$. 
The total mass including the non-relativistic part is $0.63\ M_\odot$, while the total kinetic energy reaches $ E_\mathrm{tot}=1.4\times 10^{52}$ erg, leading to a much higher energy-to-mass ratio $E_\mathrm{tot}/M_\mathrm{tot}=2.2\times 10^{52}$ erg\ $M_\odot^{-1}$ and a much shorter diffusion time scale for photons in the non-relativistic ejecta than those inferred from the associated SN component. 
It is, therefore, highly likely that the relativistic ejecta are a distinct component from the non-relativistic SN ejecta rather than continuously connected density and velocity structure. 

There are several proposed scenarios for creating ejecta moving at mildly relativistic speeds. 
Since ultra-relativistic jets are considered to be associated with the surrounding cocoon produced by the jet-star interaction, the cocoon component overwhelming the star after the jet penetration is a plausible candidate \citep{2002MNRAS.337.1349R,2005ApJ...629..903L,2013ApJ...764L..12S,2017arXiv170105198D}. 
Mildly relativistic ejecta can be produced even without the successful jet penetration, e.g., a jet choked in a massive star \citep{2011ApJ...739L..55B,2012ApJ...750...68L} or in an extended envelope attached to the star \citep{2015ApJ...807..172N}. 
In the choked jet scenarios, the energy injected by the central engine is transported by the jet through the deep interior of the star and most of the energy is dissipated in outer layers of the star, realizing relativistic ejecta with a high energy-to-mass ratio. 
In the framework of the relativistic ejecta-CSM interaction, any putative scenario for llGRBs should be able to create relativistic ejecta with the properties constrained by the light curve fitting. 
Whether the proposed scenarios can reproduce such relativistic SN ejecta should be examined thoroughly for unveiling the origin of llGRBs. 

The origin of the moderately dense CSM in the immediate vicinity of the progenitor star also remains unclear. 
The inferred value of $A_\star=25$ corresponds to a mass-loss rate of a few $10^{-4}\ M_\odot$ yr$^{-1}$ for a wind velocity of $10^3$ km s$^{-1}$, which is larger than typical mass-loss rates of galactic WN and WC stars by an order of magnitude \citep[e.g.,][]{2006A&A...457.1015H,2012A&A...540A.144S}. 
Unlike the long-lasting GRBs 060218 and 100316D, extremely large amounts of CSM ($A_\star>100$ or $\dot{M}>10^{-3}\ M_\odot$ yr$^{-1}$) are not required in this particular event. 
However, the moderately dense CSM compared to galactic Wolf-Rayet stars cannot be explained by the current standard theory of massive star evolution and stellar winds. 
On the other hand, the light curve modeling of the late-time X-ray and radio observations suggests a CSM density ($A_{\mathrm{out},\star}=0.5$) comparable to galactic Wolf-Rayet stars. 
Such centrally concentrated CSM are suggested by spectroscopic observations of type II SNe in very early stages \citep[known as ``flash spectroscopy'';][]{2014Natur.509..471G,2017NatPh..13..510Y}. 
Recent early photometric observations of type II SNe also indicate the presence of an enhanced mass-loss prior to the gravitational collapse \citep{2018NatAs.tmp..122F}. 
However, what drives the intense mass-loss and whether such dense CSM can be ubiquitously present for any types of CCSNe are still unclear. 
In addition, such intense mass-loss at the final evolutionary stage of stellar evolution may be eruptive \citep[e.g.,][]{2007ApJ...657L.105F,2013MNRAS.430.1801M}. 
Therefore, the resultant CSM may not be a simple spherical wind following the inverse square law, but a highly aspherical and/or clumpy ejecta. 
If so, it may also have a non-negligible impact on the preceding steady wind, such as the shock formation, further modifying the CSM structure. 
However, exploring a lot of possible CSM models is impractical and thus we have only modified the CSM density beyond $r=r_\mathrm{out}$ to see its influence on the theoretical light curve. 

One possible theoretical explanation is an enhanced mass-loss in late burning stages of the progenitor star by still uncertain mass-losing processes, such as wave-driven mass-loss \citep{2012MNRAS.423L..92Q,2014ApJ...780...96S}. 
Another possibility for the variation in the CSM density is an accelerating stellar wind \citep[e.g.,][]{1999isw..book.....L}, which is also suggested for early emission from type IIP SNe \citep{2017MNRAS.469L.108M,2018arXiv180207752M}. 
Stellar winds are supposed to blow slowly at the bottom and gradually accelerate up to a terminal velocity beyond the sonic point. 
Therefore, even for a constant mass-loss rate, the CSM density, which is proportional to $\dot{M}/v_\mathrm{w}$, could be enhanced in the vicinity of the stellar surface compared to a simple inverse square law. 
Although an accelerating stellar wind is a possible solution, it introduces several uncertain factors, such as, the wind velocity at the base and the velocity gradient. 
Therefore, we leave it to future work to explore appropriate wind density and velocity profiles.

We also mention that such progenitor systems may be a consequence of massive star evolution in unusual environments. 
In fact, GRBs with unusually long durations and soft late-time X-ray spectral indices, including GRBs 060218 and 100316D, appear to show large absorption column densities \citep{2015ApJ...805..159M}. 
This indicates the progenitor system of llGRBs may be closely linked with unusual environments in which the progenitor star is embedded. 
In any case, the mechanism of the enhanced mass-loss in the final evolutionary stage of massive stars is indispensable for the ultimate understanding of llGRBs and other transients powered by ejecta-CSM interaction. 

\subsection{A population of X-ray transients powered by SN ejecta-CSM interaction}
In the following, we summarize transients possibly powered by SN ejecta-CSM interaction. 

\subsubsection{Low-luminosity GRBs}\label{sec:llGRBs}
As shown in the $E_\mathrm{rad}$--$T_\mathrm{burst}$ diagram (Figure \ref{fig:diagram}), llGRBs are located in the lower right region because of their low gamma-ray luminosities. 
This clearly distinguishes llGRBs from the population of {\it Swift} GRBs with larger $E_\mathrm{iso}$ and shorter $T_{90}$. 
The prediction of the ejecta-CSM interaction model agrees with the region occupied by llGRBs. 
As \hyperlink{sms17}{SMS17} have pointed out, the locations of the previously discovered llGRBs, 980425, 060218, and 100316D, on the diagram are consistent with the region with $E_\mathrm{rel}=10^{50}$--$10^{51}$ erg and $A_\star$ of a few up to several 100.  
In addition, the newly discovered llGRB 171205A fills the gap between the fast and less energetic event (GRB 980425) and the two events with long-lasting gamma-ray emission (GRB 060218 and 100316D). 
This discovery further supports the idea that these llGRBs constitute a distinct population of transients arising from mildly relativistic SN ejecta interacting with the CSM. 
If this scenario is correct, even more events with similar properties will be detected in current and future survey missions and fill the region predicted by the model on the $E_\mathrm{rad}$--$T_\mathrm{burst}$ diagram. 

In Figure \ref{fig:diagram}, GRBs 031203, 120422A, and 161219B are located above the theoretical curves. 
In fact, GRB 031203 is often classified as a llGRB \citep[e.g.,][]{2007ApJ...657L..73G}. 
If they are actually powered by ejecta-CSM interaction, this discrepancy indicates that some modifications, e.g., aspherical ejecta, are required to account for these relatively energetic events. 
On the other hand, there are suggestions that they are intrinsically different from llGRBs. 
\cite{2012A&A...547A..82M} suggest that GRB 120422A and the associated SN 2012bz may be an intermediate case between cosmological GRBs and X-ray flashes (including events like GRB 060218). 
\cite{2012ApJ...756..190Z} found that the variable prompt gamma-ray emission of GRB 120422A is better explained by emission from a jet, rather than quasi-spherical shock emergence. 
They also suggest that a gamma-ray luminosity of $\sim 10^{48}$ erg s$^{-1}$ distinguishes GRBs driven by jets from llGRBs. 
This threshold gamma-ray luminosity is roughly consistent with the highest gamma-ray luminosity realized in our theoretical model with $E_\mathrm{rel,51}=10$. 
\cite{2017A&A...605A.107C} also classify GRB 161219B as an intermediate-luminosity GRBs. 
The presence of an intermediate class between cosmological GRBs and llGRBs and how the two populations are overlapped are still debated and thus require a larger set of examples. 

\subsubsection{XRF 080109/SN 2008D}
XRF 080109 is an X-ray flash serendipitously discovered by the {\it Swift} \citep{2008Natur.453..469S}. 
The X-ray emission was later found associated with the birth of the type Ib/c SN 2008D \citep{2008Sci...321.1185M,2009ApJ...692L..84M,2009ApJ...702..226M,2009ApJ...692.1131T}, which was supposed to be an exploding helium star with a relatively large explosion energy. 
The X-ray luminosity reaches the peak value of several $10^{43}$ erg s$^{-1}$ in the first $\sim 50$ s and then declines over the next few $100$ s. 
The total radiated energy and the burst duration are $E_\mathrm{iso}\simeq 6\times 10^{45}$ erg and $T_\mathrm{90}\simeq 470$ s \citep{2009ApJ...702..226M}, which are also plotted in Figure \ref{fig:diagram}. 
The potential similarities of XRF 080109/SN 2008D to the llGRB 060218/SN 2006aj let several authors to put forward the supernova ejecta-CSM interaction scenario for the origin of the X-ray emission.

The location of XRF 080109 on the $E_\mathrm{rad}$--$T_\mathrm{burst}$ diagram suggest that it is much less energetic than llGRBs. 
Therefore, unlike llGRBs, XRF 080109 does not appear to arise from highly energetic ejecta moving at relativistic speeds, even if it is indeed powered by ejecta-CSM interaction. 
Nevertheless, the existence of XRF 080109 probably indicates that there is an even larger population of X-ray transients associated with the birth of stripped envelope SNe. 
Unfortunately, the current detection limits of any unbiased X-ray surveys have not reached such low X-ray fluxes. 

\subsubsection{CDF-S XT1}

Recently, \cite{2017MNRAS.467.4841B} reported the detection of an X-ray transient in the Chandra Deep Field South, which is dubbed CDF-S XT1. 
Although no transient optical counterpart has been found, the most likely host galaxy was found at a photometric redshift of $z\sim 2.23$. 
Adopting the redshift, the peak $0.3$--$10$ keV X-ray luminosity reaches $2\times 10^{47}$ erg s$^{-1}$. 
The integrated X-ray energy of $9\times 10^{49}$ erg released over $\sim 10^3$ s implies a potential similarity of CDF-S XT1 to llGRBs. 
In addition, the reported photon index, $\Gamma_\mathrm{ph}=1.43^{+0.26}_{-0.15}$, of the X-ray spectrum of CDF-S XT1 is also consistent with those of llGRBs. 

Since there is no gamma-ray detection, the direct comparison of the radiated energy and the duration of CDF-S XT1 with other GRBs is not straightforward. 
However, taking the radiated energy and the duration of the X-ray emission at face values, CDF-S XT1 is located in a similar region on the $E_\mathrm{rad}$--$T_\mathrm{burst}$ diagram to llGRB 060218 and 100316D. 
Although the connection between CDF-S XT1 and llGRBs is not confirmed and warrants further investigations, this may imply that X-ray transients arising from the ejecta-CSM interaction are ubiquitously found in both nearby and high-redshift galaxies.

\subsubsection{Soft GRBs detected by MAXI}


The Monitor of All-sky X-ray Image (MAXI) on board the International Space Station (ISS) have detected 22 GRBs without any simultaneous detection by other gamma-ray satellites during the first $44$ months of its operation  (only-MAXI GRBs; \citealt{2014PASJ...66...87S}) \footnote{a more complete list including recent events is available at http://maxi.riken.jp/grbs/}. 
The non-detection by any other gamma-ray satellites indicates that only-MAXI GRBs are dominated by relatively soft X-ray photons. 
Due to the lack of successful follow-up observations, distances to only-MAXI GRBs remain unknown. 
However, as we show below, the event rate is roughly consistent with llGRBs. 

As shown by \cite{2014PASJ...66...87S}, GRBs with an average $2$--$20$ keV X-ray flux down to $\sim2\times10^{-9}$ erg s$^{-1}$ cm$^{-2}$ have been detected by MAXI. 
Assuming the lowest flux to be the detection threshold, a soft GRB with an average luminosity of $\sim 10^{46}$ erg s$^{-1}$ could be detected up to a distance of $D_\mathrm{maxi}\simeq 200$ Mpc. 
For the volumetric event rate $R_\mathrm{llgrb}$ of llGRBs estimated by several studies ($10^2$--$10^3$ Gpc$^{-3}$ yr$^{-1}$; e.g., \citealt{2006Natur.442.1014S,2006ApJ...645L.113C}), the number of events per year within the distance yields,
\begin{eqnarray}
\frac{4\pi D_\mathrm{maxi}^3R_\mathrm{llgrb}}{3}&=&
10\ \mathrm{events\ yr}^{-1}
\left(\frac{D_\mathrm{maxi}}{200\ \mathrm{Mpc}}\right)^{3}
\nonumber\\&&\ \ \ \ \times
\left(\frac{R_\mathrm{llgrb}}{300\ \mathrm{Gpc}^{-3}\ \mathrm{yr}^{-1}}\right).
\end{eqnarray}
MAXI scans the nearly entire sky every $t_\mathrm{orbit}=5500$ s, which corresponds to the single orbit of the ISS around the earth. 
Therefore, the probability of detecting a GRB with a duration $t_\mathrm{burst}$ is approximately given by $t_\mathrm{burst}/t_\mathrm{orbit}$. 
Thus, for GRB 060218-like events with burst durations of $t_\mathrm{burst}\simeq10^3$ s, the detection rate is roughly estimated to be
\begin{equation}
\frac{4\pi D_\mathrm{maxi}^3R_\mathrm{llgrb}}{3}\frac{t_\mathrm{burst}}{t_\mathrm{orbit}}=2\ \mathrm{events\ yr}^{-1}.
\end{equation}
The detection rate indeed suffers from various uncertain factors, such as, the intrinsic volumetric rate and the typical burst duration. 
Nevertheless, the estimated value roughly explains the detection rate of only-MAXI GRBs, $6$ events yr$^{-1}$ and therefore may suggest that several llGRBs have been detected by MAXI.

\section{Conclusion}\label{sec:conclusion}
In this paper, we have carried out multi-wavelength light curve modeling of the new llGRB 171205A. 
The theoretical model is based on the relativistic SN ejecta-CSM interaction scenario. 
We adopt the hydrodynamic model developed by our previous study (\hyperlink{sms17}{SMS17}), which solves the dynamical evolution of the geometrically thin shell produced by the collision between the ejecta and the CSM. 
The light curve model for the early gamma-ray emission assumes that the radiation energy produced in the shell is gradually released by radiative diffusion. 
The photospheric emission from un-shocked ejecta and non-thermal emission from the forward shock are also treated by using emission models (\hyperlink{sm18}{SM18}) combined with the hydrodynamic model. 

The broad-band emission of llGRB 171205A is successfully explained by the emission powered by the CSM interaction. 
The duration and the isotropic equivalent energy of llGRBs are also well explained by a population of relativistic SNe exploding in a relatively dense CSM, and such a population occupies a distinct region on the $E_\mathrm{rad}$--$T_\mathrm{busrt}$ diagram from cosmological GRBs. 
However, we had to introduce centrally concentrated CSM with a sudden density drop to explain the late-time X-ray and radio emission. 
Although recent observations of CCSNe provide some observational support for such CSM structure \citep[e.g.,][]{2014Natur.509..471G,2017NatPh..13..510Y}, the mechanism responsible for the enhanced mass-loss is still unclear. 
We also point out that the potential similarities of XRF 080107, CDF-S XT1, and possibly MAXI GRBs, with llGRBs may suggest that there are still a plenty of hidden or unspecified X-ray bright transients beyond the reach of current unbiased X-ray surveys. 
Future deep and/or wide-field X-ray surveys combined with intensive follow-up observations in multi-wavelengths will uncover such hidden populations and ultimately help us elucidating the still mysterious origin of highly energetic stellar explosions. 

\acknowledgments
This work made use of data supplied by the UK Swift Science Data Centre at the University of Leicester. 
AS would like to appreciate Kunihito Ioka for fruitful discussion. 
Numerical computations were in part carried out on Cray XC30 and XC50 at Center for Computational Astrophysics, National Astronomical Observatory of Japan. 
KM acknowledges support by the Japan Society for the Promotion of Science (JSPS) KAKENHI grant 17H02864, 18H05223, and 18H04585.

\software{XSPEC (v12.8; \citealt{1996ASPC..101...17A}), 
HEAsoft (v6.24; \citealt{1996ASPC..101...17A})}

\appendix
\section{Light curve calculation}\label{sec:light_curve}
In this section, we describe our method to calculate light curves of the emission from an expanding shell. 
We note that the general formula to calculate the observed luminosity of the emission from a relativistic flow has been derived \citep{1999ApJ...513..679G,1999ApJ...523..187W}. 
Here, we present the explicit formula for the observed luminosity for the sake of completeness. 

We assume that a distant observer see the emission from the shell whose thickness is negligible. 
The shell is initially located at the origin and then starts expanding. 
Thus, the shell radius is given as a function of time, $R_\mathrm{s}(t)$, with $R_\mathrm{s}(0)=0$. 

\begin{figure}
\begin{center}
\includegraphics[scale=0.50,bb=0 0 574 171]{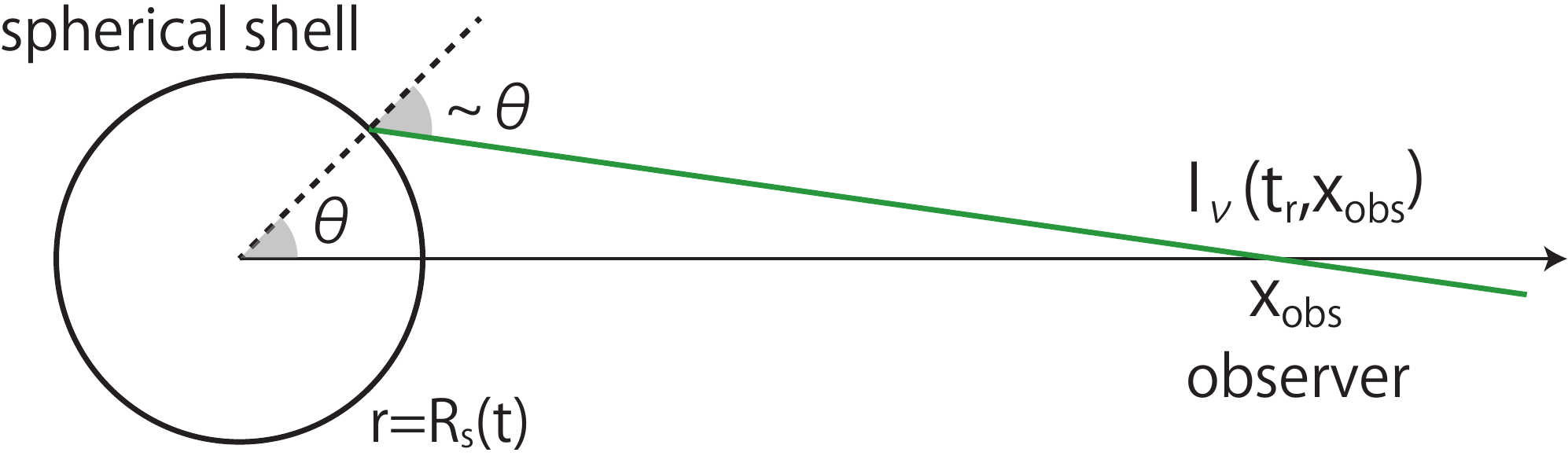}
\caption{Schematic view of the setting of the light curve calculation. }
\label{fig:shell_emission}
\end{center}
\end{figure}

Figure \ref{fig:shell_emission} schematically represents the setting of the calculation. 
The observer is standing at a distance $D$ from the origin and the corresponding position vector is denoted by $\textbf{\textit{x}}_\mathrm{obs}$. 
The distance $D$ is sufficiently larger than the size of the emitter so that the photon rays reaching the observer are almost parallel to each other. 
Therefore, for photons emitted by a point on the shell specified by the angle $\theta$, the angle between the photon ray and the radial direction is identical with $\theta$ in this limit. 
The time at which those photons at $t$ is received by the observer at $t_\mathrm{r}$, which is expressed in terms of $t$ and the direction cosine $\mu=\cos\theta$ as follows,
\begin{equation}
t_\mathrm{r}=t+\frac{D-R_\mathrm{s}(t)\mu}{c}.
\end{equation}
It is convenient to define the zero-point $t_\mathrm{obs,0}$ of the observer time $t_\mathrm{obs}$ as the time at which the photon emitted from the origin $r=0$ at $t=0$ reaches the observer, $t_\mathrm{obs,0}=D/c$. 
Thus, the observer time is given by 
\begin{equation}
t_\mathrm{obs}=t_\mathrm{r}-t_\mathrm{obs,0}=t-\frac{R_\mathrm{s}(t)}{c}\mu
\end{equation}

The emissivity for a frequency $\nu$ at a position $\textbf{\textit{x}}$ and time $t$ in the rest frame of the observer is given by
\begin{equation}
j_\nu(t,\textbf{\textit{x}})=\frac{d\dot{E}(t,\bar{\nu})/d\nu}{4\pi\Omega_\mathrm{em}\Gamma_\mathrm{s}^2[1-\beta_\mathrm{s}(t)
\mu]^2R_\mathrm{s}(t)^2}
\delta(|\textbf{\textit{x}}|-R_\mathrm{s}),
\end{equation}
where $\Omega_\mathrm{em}$ is the solid angle into which photons are emitted. 
The energy loss rate per unit frequency in the rest frame of a segment of the shell is denoted by $d\dot{E}(t,\nu)/d\nu$. 
The solid angle is $\Omega_\mathrm{em}=4\pi$ for an optically thin shell, while $\Omega_\mathrm{em}=2\pi$ for an optically thick shell, because photons are emitted only for the outer region of the shell in the optically thick case. 
The frequency $\bar{\nu}$ in the rest frame of the shell is related to the observer-frame frequency by
\begin{equation}
\bar{\nu}=\Gamma_\mathrm{s}[1-\beta_\mathrm{s}(t)\mu]\nu
\label{eq:nu_bar}
\end{equation}
The integration of the radiative transfer equation along a direction vector $\textbf{\textit{l}}$ leads to the intensity at the observer at the position $\textbf{\textit{x}}$ and the time $t=t_\mathrm{r}$,
\begin{equation}
I_\nu(t_\mathrm{r},\textbf{\textit{x}}_\mathrm{obs})=c\int_0^{t_\mathrm{r}}j_\nu(t,\textbf{\textit{x}}_\mathrm{obs}-c(t_\mathrm{r}-t)\textbf{\textit{l}})dt,
\end{equation} 
Integrating the intensity multiplied by the direction cosine $\textbf{\textit{x}}_\mathrm{obs}\cdot\textbf{\textit{l}}/D$, one obtains the observed flux,
\begin{equation}
F_{\nu,\mathrm{obs}}(t_\mathrm{obs})=\frac{c}{2\Omega_\mathrm{em} D^2}
\int^{t_\mathrm{obs}+t_\mathrm{obs,0}}_0\frac{d\dot{E}(t,\bar{\nu})/d\nu}{\Gamma^2[1-\beta_\mathrm{s}(t)\mu]^2R_\mathrm{s}(t)}dt,
\end{equation}
where we have assumed $D\gg R_\mathrm{s}(t)$. 
The corresponding luminosity per unit frequency is given by
\begin{equation}
L_{\nu,\mathrm{obs}}(t_\mathrm{obs})=\frac{2\pi c}{\Omega_\mathrm{em}}
\int^{t_\mathrm{obs}+t_\mathrm{obs,0}}_0\frac{d\dot{E}(t,\bar{\nu})/d\nu}{\Gamma^2[1-\beta_\mathrm{s}(t)\mu]^2R_\mathrm{s}(t)}dt,
\end{equation}
where the direction cosine $\mu$ is given by
\begin{equation}
\mu=\frac{c}{R_\mathrm{s}(t)}(t-t_\mathrm{obs})
\label{eq:mu}
\end{equation}
The range of the integration with respect to $t$ is restricted so that $\mu$ takes physically meaningful values. 
For optically thin and thick cases, the direction cosine should be in the following ranges;
\begin{equation}
-1\leq \mu \leq 1
\end{equation}
for an optically thin shell, and
\begin{equation}
\beta_\mathrm{s}(t)\leq \mu\leq 1,
\end{equation}
for an optically thick shell. 
The latter condition $\beta_\mathrm{s}(t)\leq \mu$ for the optically thick case guarantees that photons once emitted by the shell do not intersect with the shell again. 

The bolometric luminosity is obtained by integrating the luminosity per unit frequency with respect to the frequency $\nu$ and found to be
\begin{equation}
L_\mathrm{obs}(t_\mathrm{obs})=\frac{2\pi c}{\Omega_\mathrm{em}}
\int^{t_\mathrm{obs}+t_\mathrm{obs,0}}_0\frac{\dot{E}(t)}{\Gamma^3[1-\beta_\mathrm{s}(t)\mu]^3R_\mathrm{s}(t)}dt,
\end{equation}
where $\dot{E}(t)$ is the frequency-integrated energy loss rate given by
\begin{equation}
\dot{E}(t)=\int \frac{d\dot{E}(t,\bar{\nu})}{d\nu}d\bar{\nu}. 
\end{equation}

\section{Time-sliced XRT spectra}\label{sec:xrt}
In this section, we describe our procedure to obtain the XRT spectra plotted in Figure \ref{fig:sed}. 
UK Swift Science Data Centre provides XRT data for requested time intervals \citep[see][for detail]{2009MNRAS.397.1177E}. 
We have obtained XRT data for the three time intervals, from $t_\mathrm{obs}=3\times 10^3$ to $10^4$ s, from $t_\mathrm{obs}=10^4$ to $3\times 10^4$ s, and from $t_\mathrm{obs}=10^5$ to $2\times 10^5$ s. 
Then, we analyzed XRT data for each interval in the following way. 

We use the X-ray spectral fitting package Xspec\footnote{https://heasarc.gsfc.nasa.gov/docs/software/lheasoft/}, developed and distributed by the High Energy Astrophysics Science Archive Research Center (HEASARC) at NASA. 
The observed X-ray spectrum is fitted by an absorbed power-law function in a standard way similar to \cite{2009MNRAS.397.1177E}. 
Specifically, we made use of the model {\it phabs*zphabs*powerlaw} with a fixed hydrogen column density $N_H=5.89\times 10^{20}$ cm$^{-2}$ for the first absorption component corresponding to the Galactic absorption. 
For the second absorption component {\it zphabs}, which is the local one associated with GRB 171205A, we fixed the redshift of $0.0368$. 
The results of the fitting are summarized in Table \ref{table:fitting}. 
We note that our analysis gave similar results to those automatically created by UK Swift Science Data Centre. 

\begin{figure*}
\begin{center}
\includegraphics[scale=0.55,bb=0 0 864 576]{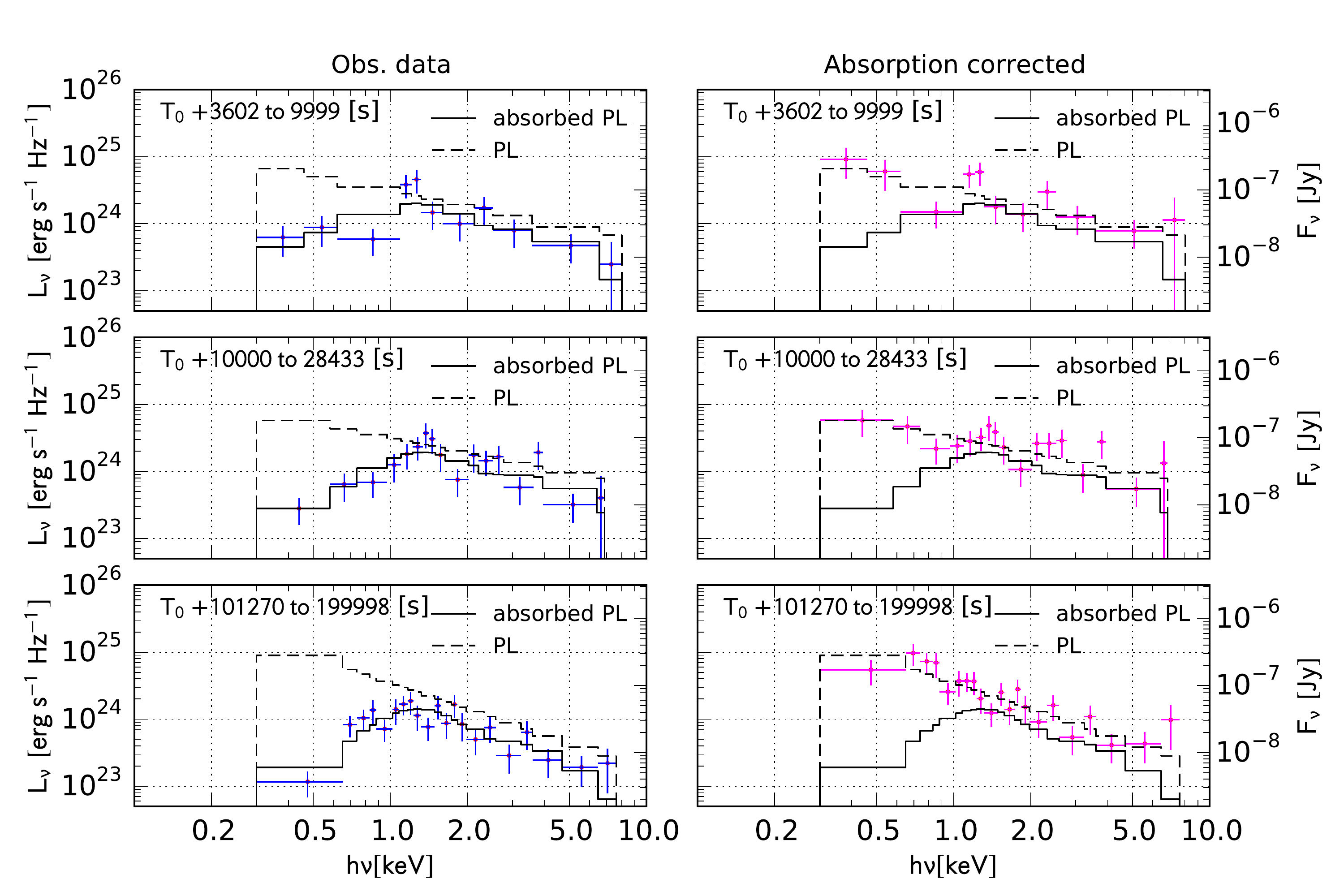}
\caption{XRT spectra and model fitting results. 
In the left and right panels, observational and absorption corrected data are plotted for different time intervals. 
In all panels, the best-fit absorbed power-law function is plotted as a solid line and the corresponding unabsorbed power-law function is shown as a dashed line. }
\label{fig:xrt}
\end{center}
\end{figure*}

The left panels in Figure \ref{fig:xrt} show the observational data and the fitting results for different time intervals. 
The absorbed single power-law function (solid line) well reproduces the observational data. 
We also plotted the corresponding unabsorbed power-law function as a dashed line in each panel. 
The ratio of the absorbed to unabsorbed power-law function represents the extent of the absorption and used to correct the observed spectra. 
The spectra shown in the right panels of Figure \ref{fig:xrt} are produced by the absorption correction and same as XRT spectra shown in Figure \ref{fig:sed}. 

\begin{table*}
\begin{center}
  \caption{SPECTRAL FITTING RESULTS}
\begin{tabular}{ccccc}
\hline\hline
time intervel&$T_0+3602\ \mathrm{to}\ 9999\ \mathrm{s}$&$T_0+10000\ \mathrm{to}\ 28433\ \mathrm{s}$&$T_0+101270\ \mathrm{to}\ 199998\ \mathrm{s}$\\
\hline
Galactic $N_H$ (fixed) [$10^{22}\mathrm{cm}^{-2}$]&0.0589&0.0589&0.0589\\
intrinsic $N_H$ [$10^{22}\mathrm{cm}^{-2}$]&$0.0\pm0.0141$&$0.0823\pm0.0761$&$0.188\pm0.0838$\\
photon index&$1.77\pm0.353$&$1.73\pm0.309$&$2.28\pm0.277$\\
unabsorbed $0.3$--$10$ keV flux [$10^{-12}$ erg cm$^{-2}$ s$^{-1}$]&$0.967^{+0.177}_{-0.149}$&$1.01^{+0.141}_{-0.124}$&$0.845^{+0.176}_{-0.145}$\\
model $0.3$--$10$ keV flux [$10^{-12}$ erg cm$^{-2}$ s$^{-1}$]&$0.857$&$0.823$&$0.468$\\
C-statistics (d.o.f.)&$48.46(48)$&$76.00(78)$&$76.31(100)$\\
\hline\hline
\end{tabular}
\label{table:fitting}
\end{center}
\end{table*}

\begin{thebibliography}{}
\bibitem[Afsariardchi \& Matzner(2018)]{2018ApJ...856..146A} Afsariardchi, N., \& Matzner, C.~D.\ 2018, \apj, 856, 146 
\bibitem[Arnaud(1996)]{1996ASPC..101...17A} Arnaud, K.~A.\ 1996, Astronomical Data Analysis Software and Systems V, 101, 17 
\bibitem[Barthelmy et al.(2017)]{2017GCN.22184....1B} Barthelmy, S.~D., Cummings, J.~R., D'Elia, V., et al.\ 2017, GRB Coordinates Network, Circular Service, No.~22184, \#1 (2017), 22184, 1 
\bibitem[Bauer et al.(2017)]{2017MNRAS.467.4841B} Bauer, F.~E., Treister, E., Schawinski, K., et al.\ 2017, \mnras, 467, 4841
\bibitem[Barniol Duran et al.(2015)]{2015MNRAS.448..417B} Barniol Duran, R., Nakar, E., Piran, T., \& Sari, R.\ 2015, \mnras, 448, 417 
\bibitem[Breeveld et al.(2011)]{2011AIPC.1358..373B} Breeveld, A.~A., Landsman, W., Holland, S.~T., et al.\ 2011, American Institute of Physics Conference Series, 1358, 373 
\bibitem[Bromberg et al.(2011)]{2011ApJ...739L..55B} Bromberg, O., Nakar, E., \& Piran, T.\ 2011, \apjl, 739, L55 
\bibitem[Brown et al.(2010)]{2010ApJ...721.1608B} Brown, P.~J., Roming, P.~W.~A., Milne, P., et al.\ 2010, \apj, 721, 1608 
\bibitem[Bufano et al.(2012)]{2012ApJ...753...67B} Bufano, F., Pian, E., Sollerman, J., et al.\ 2012, \apj, 753, 67
\bibitem[Campana et al.(2006)]{2006Natur.442.1008C} Campana, S., Mangano, V., Blustin, A.~J., et al.\ 2006, \nat, 442, 1008 
\bibitem[Cano et al.(2011)]{2011ApJ...740...41C} Cano, Z., Bersier, D., Guidorzi, C., et al.\ 2011, \apj, 740, 41 
\bibitem[Cano et al.(2017a)]{2017A&A...605A.107C} Cano, Z., Izzo, L., de Ugarte Postigo, A., et al.\ 2017, \aap, 605, A107 
\bibitem[Cano et al.(2017b)]{2017AdAst2017E...5C} Cano, Z., Wang, S.-Q., Dai, Z.-G., \& Wu, X.-F.\ 2017, Advances in Astronomy, 2017, 8929054 
\bibitem[Cardelli et al.(1989)]{1989ApJ...345..245C} Cardelli, J.~A., Clayton, G.~C., \& Mathis, J.~S.\ 1989, \apj, 345, 245 
\bibitem[Chakraborti et al.(2015)]{2015ApJ...805..187C} Chakraborti, S., Soderberg, A., Chomiuk, L., et al.\ 2015, \apj, 805, 187 
\bibitem[Chevalier(1982a)]{1982ApJ...258..790C} Chevalier, R.~A.\ 1982, \apj, 258, 790
\bibitem[Chevalier(1982b)]{1982ApJ...259..302C} Chevalier, R.~A.\ 1982, \apj, 259, 302 
\bibitem[Chevalier \& Fransson(2016)]{2016arXiv161207459C} Chevalier, R.~A., \& Fransson, C.\ 2016, arXiv:1612.07459 
\bibitem[Cobb et al.(2006)]{2006ApJ...645L.113C} Cobb, B.~E., Bailyn, C.~D., van Dokkum, P.~G., \& Natarajan, P.\ 2006, \apjl, 645, L113
\bibitem[Dado \& Dar(2017)]{2017arXiv171209319D} Dado, S., \& Dar, A.\ 2017, arXiv:1712.09319 
\bibitem[De Colle et al.(2017)]{2017arXiv170105198D} De Colle, F., Lu, W., Kumar, P., Ramirez-Ruiz, E., \& Smoot, G.\ 2017, arXiv:1701.05198 
\bibitem[D'Elia et al.(2017)]{2017GCN.22177....1D} D'Elia, V., D'Ai, A., Lien, A.~Y., \& Sbarufatti, B.\ 2017, GRB Coordinates Network, Circular Service, No.~22177, \#1 (2017), 22177, 1 
\bibitem[D'Elia et al.(2018)]{delia2018} D'Elia, V., Campana, S., D'A{\`i}, A., et al.\ 2018, \aap in press (arXiv:1810.03339)
\bibitem[de Ugarte Postigo et al.(2017a)]{2017GCN.22204....1D} de Ugarte Postigo, A., Izzo, L., Kann, D.~A., et al.\ 2017, GRB Coordinates Network, Circular Service, No.~22204, \#1 (2017), 22204, 1 
\bibitem[de Ugarte Postigo et al.(2017b)]{2017GCN.22187....1D} de Ugarte Postigo, A., Schulze, S., Bremer, M., et al.\ 2017, GRB Coordinates Network, Circular Service, No.~22187, \#1 (2017), 22187, 1 
\bibitem[Evans et al.(2007)]{2007A&A...469..379E} Evans, P.~A., Beardmore, A.~P., Page, K.~L., et al.\ 2007, \aap, 469, 379 
\bibitem[Evans et al.(2009)]{2009MNRAS.397.1177E} Evans, P.~A., Beardmore, A.~P., Page, K.~L., et al.\ 2009, \mnras, 397, 1177
\bibitem[Foley et al.(2007)]{2007ApJ...657L.105F} Foley, R.~J., Smith, N., Ganeshalingam, M., et al.\ 2007, \apjl, 657, L105 
\bibitem[F{\"o}rster et al.(2018)]{2018NatAs.tmp..122F} F{\"o}rster, F., Moriya, T.~J., Maureira, J.~C., et al.\ 2018, Nature Astronomy, 2, 808
\bibitem[Galama et al.(1998)]{1998Natur.395..670G} Galama, T.~J., Vreeswijk, P.~M., van Paradijs, J., et al.\ 1998, \nat, 395, 670 
\bibitem[Gal-Yam et al.(2014)]{2014Natur.509..471G} Gal-Yam, A., Arcavi, I., Ofek, E.~O., et al.\ 2014, \nat, 509, 471 
\bibitem[Gehrels et al.(2004)]{2004ApJ...611.1005G} Gehrels, N., Chincarini, G., Giommi, P., et al.\ 2004, \apj, 611, 1005 
\bibitem[Granot et al.(1999)]{1999ApJ...513..679G} Granot, J., Piran, T., \& Sari, R.\ 1999, \apj, 513, 679 
\bibitem[Granot \& Sari(2002)]{2002ApJ...568..820G} Granot, J., \& Sari, R.\ 2002, \apj, 568, 820 
\bibitem[Guetta \& Della Valle(2007)]{2007ApJ...657L..73G} Guetta, D., \& Della Valle, M.\ 2007, \apjl, 657, L73 
\bibitem[Hamann et al.(2006)]{2006A&A...457.1015H} Hamann, W.-R., Gr{\"a}fener, G., \& Liermann, A.\ 2006, \aap, 457, 1015
\bibitem[Hjorth \& Bloom(2012)]{2012grbu.book..169H} Hjorth, J., \& Bloom, J.~S.\ 2012, Chapter 9 in ''Gamma-Ray Bursts'', Cambridge Astrophysics Series 51,  eds.~C.~Kouveliotou, R.~A.~M.~J.~Wijers and S.~Woosley, Cambridge University Press (Cambridge), p.~169-190, 169
\bibitem[Irwin \& Chevalier(2016)]{2016MNRAS.460.1680I} Irwin, C.~M., \& Chevalier, R.~A.\ 2016, \mnras, 460, 1680 
\bibitem[Izzo et al.(2017)]{2017GCN.22180....1I} Izzo, L., Selsing, J., Japelj, J., et al.\ 2017, GRB Coordinates Network, Circular Service, No.~22180, \#1 (2017), 22180, 1 
\bibitem[Kennea et al.(2017)]{2017GCN.22183....1K} Kennea, J.~A., Sbarufatti, B., Burrows, D.N., et al.\ 2017, GRB Coordinates Network, Circular Service, No.~22183, \#1 (2017), 22183, 1 
\bibitem[Kulkarni et al.(1998)]{1998Natur.395..663K} Kulkarni, S.~R., Frail, D.~A., Wieringa, M.~H., et al.\ 1998, \nat, 395, 663 
\bibitem[Lamers \& Cassinelli(1999)]{1999isw..book.....L} Lamers, H.~J.~G.~L.~M., \& Cassinelli, J.~P.\ 1999, Introduction to Stellar Winds, by Henny J.~G.~L.~M.~Lamers and Joseph P.~Cassinelli, pp.~452.~ISBN 0521593980.~Cambridge, UK: Cambridge University Press, June 1999., 452 
\bibitem[Laskar et al.(2017)]{2017GCN.22216....1L} Laskar, T., Coppejans, D.~L., Margutti, R., \& Alexander, K.~D.\ 2017, GRB Coordinates Network, Circular Service, No.~22216, \#1 (2017), 22216, 1
\bibitem[Lazzati \& Begelman(2005)]{2005ApJ...629..903L} Lazzati, D., \& Begelman, M.~C.\ 2005, \apj, 629, 903
\bibitem[Lazzati et al.(2012)]{2012ApJ...750...68L} Lazzati, D., Morsony, B.~J., Blackwell, C.~H., \& Begelman, M.~C.\ 2012, \apj, 750, 68 
\bibitem[Li(2007)]{2007MNRAS.375..240L} Li, L.-X.\ 2007, \mnras, 375, 240
\bibitem[Li \& Chevalier(1999)]{1999ApJ...526..716L} Li, Z.-Y., \& Chevalier, R.~A.\ 1999, \apj, 526, 716
\bibitem[Liang et al.(2007)]{2007ApJ...662.1111L} Liang, E., Zhang, B., Virgili, F., \& Dai, Z.~G.\ 2007, \apj, 662, 1111 
\bibitem[Lien et al.(2016)]{2016ApJ...829....7L} Lien, A., Sakamoto, T., Barthelmy, S.~D., et al.\ 2016, \apj, 829, 7
\bibitem[Malesani et al.(2009)]{2009ApJ...692L..84M} Malesani, D., Fynbo, J.~P.~U., Hjorth, J., et al.\ 2009, \apjl, 692, L84
\bibitem[Margutti et al.(2015)]{2015ApJ...805..159M} Margutti, R., Guidorzi, C., Lazzati, D., et al.\ 2015, \apj, 805, 159 
\bibitem[Matzner et al.(2013)]{2013ApJ...779...60M} Matzner, C.~D., Levin, Y., \& Ro, S.\ 2013, \apj, 779, 60 
\bibitem[Matzner \& McKee(1999)]{1999ApJ...510..379M} Matzner, C.~D., \& McKee, C.~F.\ 1999, \apj, 510, 379 
\bibitem[Mauerhan et al.(2013)]{2013MNRAS.430.1801M} Mauerhan, J.~C., Smith, N., Filippenko, A.~V., et al.\ 2013, \mnras, 430, 1801 
\bibitem[Mazzali et al.(2006)]{2006Natur.442.1018M} Mazzali, P.~A., Deng, J., Nomoto, K., et al.\ 2006, \nat, 442, 1018 
\bibitem[Mazzali et al.(2008)]{2008Sci...321.1185M} Mazzali, P.~A., Valenti, S., Della Valle, M., et al.\ 2008, Science, 321, 1185
\bibitem[Melandri et al.(2012)]{2012A&A...547A..82M} Melandri, A., Pian, E., Ferrero, P., et al.\ 2012, \aap, 547, A82 
\bibitem[Milisavljevic et al.(2015)]{2015ApJ...799...51M} Milisavljevic, D., Margutti, R., Parrent, J.~T., et al.\ 2015, \apj, 799, 51 
\bibitem[Modjaz et al.(2006)]{2006ApJ...645L..21M} Modjaz, M., Stanek, K.~Z., Garnavich, P.~M., et al.\ 2006, \apjl, 645, L21 
\bibitem[Modjaz et al.(2009)]{2009ApJ...702..226M} Modjaz, M., Li, W., Butler, N., et al.\ 2009, \apj, 702, 226 
\bibitem[Moriya et al.(2018)]{2018arXiv180207752M} Moriya, T.~J., F{\"o}rster, F., Yoon, S.-C., Gr{\"a}fener, G., \& Blinnikov, S.~I.\ 2018, arXiv:1802.07752 
\bibitem[Moriya et al.(2017)]{2017MNRAS.469L.108M} Moriya, T.~J., Yoon, S.-C., Gr{\"a}fener, G., \& Blinnikov, S.~I.\ 2017, \mnras, 469, L108 
\bibitem[Nakar(2015)]{2015ApJ...807..172N} Nakar, E.\ 2015, \apj, 807, 172 
\bibitem[Nakar \& Sari(2012)]{2012ApJ...747...88N} Nakar, E., \& Sari, R.\ 2012, \apj, 747, 88
\bibitem[Nakauchi et al.(2015)]{2015ApJ...805..164N} Nakauchi, D., Kashiyama, K., Nagakura, H., Suwa, Y., \& Nakamura, T.\ 2015, \apj, 805, 164 
\bibitem[Ohtani et al.(2018)]{2018ApJ...853...52O} Ohtani, Y., Suzuki, A., Shigeyama, T., \& Tanaka, M.\ 2018, \apj, 853, 52 
\bibitem[Olivares E.~et al.(2012)]{2012A&A...539A..76O} Olivares E., F., Greiner, J., Schady, P., et al.\ 2012, \aap, 539, A76 
\bibitem[Perley et al.(2017)]{2017GCN.22252....1P} Perley, D.~A., Schulze, S., \& de Ugarte Postigo, A.\ 2017, GRB Coordinates Network, Circular Service, No.~22252, \#1 (2017), 22252, 1 
\bibitem[Pian et al.(2006)]{2006Natur.442.1011P} Pian, E., Mazzali, P.~A., Masetti, N., et al.\ 2006, \nat, 442, 1011 
\bibitem[Poole et al.(2008)]{2008MNRAS.383..627P} Poole, T.~S., Breeveld, A.~A., Page, M.~J., et al.\ 2008, \mnras, 383, 627 
\bibitem[Quataert \& Shiode(2012)]{2012MNRAS.423L..92Q} Quataert, E., \& Shiode, J.\ 2012, \mnras, 423, L92 
\bibitem[Ramirez-Ruiz et al.(2002)]{2002MNRAS.337.1349R} Ramirez-Ruiz, E., Celotti, A., \& Rees, M.~J.\ 2002, \mnras, 337, 1349 
\bibitem[Rybicki \& Lightman(1979)]{1979rpa..book.....R} Rybicki, G.~B., \& Lightman, A.~P.\ 1979, New York, Wiley-Interscience, 1979.~393 p., 
\bibitem[Sander et al.(2012)]{2012A&A...540A.144S} Sander, A., Hamann, W.-R., \& Todt, H.\ 2012, \aap, 540, A144 
\bibitem[Sari et al.(1998)]{1998ApJ...497L..17S} Sari, R., Piran, T., \& Narayan, R.\ 1998, \apjl, 497, L17 
\bibitem[Sari \& Esin(2001)]{2001ApJ...548..787S} Sari, R., \& Esin, A.~A.\ 2001, \apj, 548, 787 
\bibitem[Serino et al.(2014)]{2014PASJ...66...87S} Serino, M., Sakamoto, T., Kawai, N., et al.\ 2014, \pasj, 66, 87 
\bibitem[Shiode \& Quataert(2014)]{2014ApJ...780...96S} Shiode, J.~H., \& Quataert, E.\ 2014, \apj, 780, 96 
\bibitem[Siegel et al.(2017)]{2017GCN.22202....1S} Siegel, M.~H., Brown, P.~J., Emery, S.~W.~K., \& D'Elia, V.\ 2017, GRB Coordinates Network, Circular Service, No.~22202, \#1 (2017), 22202, 1 
\bibitem[Soderberg et al.(2008)]{2008Natur.453..469S} Soderberg, A.~M., Berger, E., Page, K.~L., et al.\ 2008, \nat, 453, 469
\bibitem[Soderberg et al.(2010)]{2010Natur.463..513S} Soderberg, A.~M., Chakraborti, S., Pignata, G., et al.\ 2010, \nat, 463, 513 

\bibitem[Soderberg et al.(2006)]{2006Natur.442.1014S} Soderberg, A.~M., Kulkarni, S.~R., Nakar, E., et al.\ 2006, \nat, 442, 1014
\bibitem[Starling et al.(2011)]{2011MNRAS.411.2792S} Starling, R.~L.~C., Wiersema, K., Levan, A.~J., et al.\ 2011, \mnras, 411, 2792  
\bibitem[Suzuki et al.(2016)]{2016ApJ...825...92S} Suzuki, A., Maeda, K., \& Shigeyama, T.\ 2016, \apj, 825, 92
\bibitem[Suzuki et al.(2017)]{2017ApJ...834...32S} Suzuki, A., Maeda, K., \& Shigeyama, T.\ 2017, \apj, 834, 32 \hypertarget{sms17}{(SMS17)}
\bibitem[Suzuki \& Maeda(2018)]{2018MNRAS.478..110S} Suzuki, A., \& Maeda, K.\ 2018, \mnras, 478, 110 
 \hypertarget{sm18}{(SM18)}
\bibitem[Suzuki \& Shigeyama(2010)]{2010ApJ...717L.154S} Suzuki, A., \& Shigeyama, T.\ 2010, \apjl, 717, L154
\bibitem[Suzuki \& Shigeyama(2013)]{2013ApJ...764L..12S} Suzuki, A., \& Shigeyama, T.\ 2013, \apjl, 764, L12
\bibitem[Tan et al.(2001)]{2001ApJ...551..946T} Tan, J.~C., Matzner, C.~D., \& McKee, C.~F.\ 2001, \apj, 551, 946 
\bibitem[Tanaka et al.(2009)]{2009ApJ...692.1131T} Tanaka, M., Tominaga, N., Nomoto, K., et al.\ 2009, \apj, 692, 1131 
\bibitem[Toma et al.(2007)]{2007ApJ...659.1420T} Toma, K., Ioka, K., Sakamoto, T., \& Nakamura, T.\ 2007, \apj, 659, 1420
\bibitem[Wang et al.(2018)]{2018arXiv181003250W} Wang, J., Zhu, Z.~P., Xu, D., et al.\ 2018, arXiv:1810.03250 
\bibitem[Wang et al.(2007)]{2007ApJ...664.1026W} Wang, X.-Y., Li, Z., Waxman, E., \& M{\'e}sz{\'a}ros, P.\ 2007, \apj, 664, 1026
\bibitem[Waxman et al.(2007)]{2007ApJ...667..351W} Waxman, E., M{\'e}sz{\'a}ros, P., \& Campana, S.\ 2007, \apj, 667, 351 
\bibitem[Weiler et al.(2002)]{2002ARA&A..40..387W} Weiler, K.~W., Panagia, N., Montes, M.~J., \& Sramek, R.~A.\ 2002, \araa, 40, 387 
\bibitem[Woods \& Loeb(1999)]{1999ApJ...523..187W} Woods, E., \& Loeb, A.\ 1999, \apj, 523, 187 
\bibitem[Woosley et al.(1999)]{1999ApJ...516..788W} Woosley, S.~E., Eastman, R.~G., \& Schmidt, B.~P.\ 1999, \apj, 516, 788 
\bibitem[Woosley \& Bloom(2006)]{2006ARA&A..44..507W} Woosley, S.~E., \& Bloom, J.~S.\ 2006, \araa, 44, 507 
\bibitem[Yaron et al.(2017)]{2017NatPh..13..510Y} Yaron, O., Perley, D.~A., Gal-Yam, A., et al.\ 2017, Nature Physics, 13, 510 
\bibitem[Zhang et al.(2012)]{2012ApJ...756..190Z} Zhang, B.-B., Fan, Y.-Z., Shen, R.-F., et al.\ 2012, \apj, 756, 190 
\end{thebibliography}
\end{document}